\documentclass{emulateapj}
\usepackage{apjfonts}
\newcommand{\figexpand}{\epsscale{1.15}}
\newcommand{\plotter}{\plotone}

\newcommand{\etal}{et al.}
\newcommand{\mbh}{M_{\rm BH}}
\newcommand{\mstar}{M_{\ast}}
\newcommand{\lstar}{L_{\ast}}

\newcommand{\fgas}{f_{\rm gas}}

\newcommand{\msun}{M_{\sun}}
\newcommand{\tH}{t_{\rm H}}
\newcommand{\tmerger}{t_{\rm merger}}

\newcommand{\mhalo}{M_{\rm halo}}
\newcommand{\mgal}{M_{\rm gal}}

\newcommand{\paperone}{Paper \textrm{I}}
\newcommand{\papertwo}{Paper \textrm{II}}

\shorttitle{Co-Evolution of Quasars, Black Holes, and Galaxies \textrm{II}}
\shortauthors{Hopkins \etal}
\slugcomment{Submitted to ApJ, June 8, 2007}
\begin{document}

\title{A Cosmological Framework for the Co-Evolution of Quasars,
Supermassive Black Holes, and Elliptical Galaxies: \textrm{II}. Formation of Red Ellipticals}
\author{Philip F. Hopkins\altaffilmark{1}, 
Thomas J. Cox\altaffilmark{1}, 
Du{\v s}an Kere{\v s}\altaffilmark{1}, 
\&\ Lars Hernquist\altaffilmark{1}
}
\altaffiltext{1}{Harvard-Smithsonian Center for Astrophysics, 
60 Garden Street, Cambridge, MA 02138}

\begin{abstract}
We develop and test a model for the cosmological role of mergers 
in the formation and quenching of red, early-type galaxies. 
By combining theoretically well-constrained 
halo and subhalo mass functions as a function of redshift and 
environment with empirical halo occupation models, we predict the distribution of 
mergers as a function of redshift, environment, and physical galaxy properties. 
Making the simple ansatz that star formation is quenched after a gas-rich, 
spheroid-forming major merger, we demonstrate that this naturally 
predicts the turnover in the efficiency of star formation and baryon 
fractions in galaxies at $\sim\lstar$ (without any parameters tuned to 
this value), as well as the 
observed mass functions and mass density of red galaxies as a function of 
redshift, the formation times of early-type galaxies as a function of mass, and 
the fraction of quenched galaxies as a function of galaxy and halo mass, environment, 
and redshift. Comparing to a variety of semi-analytic 
models in which quenching is primarily driven 
by halo mass considerations or secular/disk instabilities, 
we demonstrate that our model makes unique and robust qualitative predictions 
for a number of observables, including the bivariate red fraction as a function of 
galaxy and halo mass, the density of passive galaxies at high redshifts, 
the emergence/evolution of the color-morphology-density relations 
at high redshift, and the fraction of disky/boxy (or cusp/core) spheroids  
as a function of mass. In each case, the observations favor a model 
in which some mechanism quenches future star formation after a major merger 
builds a massive spheroid. Models where quenching is dominated by a halo mass 
threshold fail to match the behavior of the bivariate red fractions, predict 
too low a density of passive galaxies at high redshift, and overpredict 
by an order of magnitude the mass of the transition from disky to boxy ellipticals. 
Models driven by secular disk instabilities also qualitatively disagree with the 
bivariate red fractions, fail to predict the observed evolution in the 
color-density relations, and predict order-of-magnitude 
incorrect distributions of kinematic types in early-type galaxies. 
We make specific predictions for how future observations, 
for example quantifying the red fraction as a function of galaxy mass, halo mass, 
environment, or redshift, 
can break the degeneracies between a number of different assumptions 
adopted in present galaxy formation models.
We discuss a variety of physical possibilities 
for this quenching, and propose a mixed scenario in which traditional quenching 
in hot, quasi-static massive halos is supplemented by the strong shocks and 
feedback energy input associated with a major merger (e.g.\ tidal shocks, 
starburst-driven winds, and quasar feedback), which 
temporarily suppress cooling and establish the conditions of 
a dynamically hot halo in the central regions of the host, even in low mass halos (below 
the traditional threshold for accretion shocks). 
\end{abstract}


\section{Introduction}
\label{sec:intro}

Recent, large galaxy 
surveys such as SDSS, 2dFGRS, COMBO-17, and DEEP 
have demonstrated that the local distribution of galaxies is 
bimodal with respect to a number of physical properties, including 
color, morphology, star formation, concentration, and surface brightness, 
\citep[e.g.][]{strateva:color.bimodality}, 
and that this bimodality extends at least to moderate
redshifts, $z\sim1.5$ \citep[e.g.,][]{bell:combo17.lfs,willmer:deep2.lfs} with a 
significant 
population of massive, red, passively evolving galaxies at even higher redshifts
\citep{labbe05:drgs,kriek:drg.seds}. 
The massive red galaxies in this bimodal
distribution correspond to traditional spheroids, with high surface brightness and 
concentration \citep{kauffmann:bimodality}, 
with little continuing star formation since their formation 
at early times \citep{trager:ages}. Understanding the formation, and in particular the 
turning off or ``quenching'' of star formation on the red sequence, is therefore 
of fundamental importance to understanding the origin of galaxies.

Hierarchical theories of galaxy formation and evolution indicate that
large systems are built up over time through the merger of smaller
progenitors, and galaxy interactions in the local Universe motivate the
``merger hypothesis'' \citep{toomre72,toomre77}, according
to which collisions between spiral galaxies produce the massive
ellipticals observed at present times. 

Observations increasingly support the notion that galaxy mergers produce 
starbursts and structure ellipticals. The 
most intense starbursts, ultraluminous infrared galaxies (ULIRGs),
are always associated with mergers \citep[e.g.][]{sanders96:ulirgs.mergers}, with 
dense gas in their centers providing material to feed black hole (BH) 
growth and to boost the
concentration and central phase space density to match those of
ellipticals \citep{hernquist:phasespace,robertson:fp}. 
Likewise, observations of individual merging systems and 
gas-rich merger remnants
\citep[e.g.,][]{LakeDressler86,Doyon94,ShierFischer98,James99}, 
as well as post-starburst (E+A/K+A) galaxies 
\citep{goto:e+a.merger.connection}, have shown that their kinematic and
photometric properties, including velocity dispersions,
concentrations, stellar masses, light profiles, and phase space
densities, are consistent with their eventual evolution into typical
$\sim L_{\ast}$ elliptical galaxies. The correlations
obeyed by these mergers and remnants
\citep[e.g.,][]{Genzel01,rothberg.joseph:kinematics,rothberg.joseph:rotation} 
are similar to e.g.\ the observed
fundamental plane and Kormendy relations for relaxed ellipticals, and
consistent with evolution onto these relations as their stellar populations
age. This is further supported by the ubiquitous presence of fine structures such as shells, ripples,
and tidal plumes in ellipticals \citep[e.g.][]{schweizerseitzer92,schweizer96}, 
which are signatures of mergers 
\citep[e.g.][]{quinn.84,hernquist.quinn.87,hernquist.spergel.92}, 
and the clustering and mass density of ellipticals, consistent with 
passive evolution after formation in mergers \citep{hopkins:clustering}. 

Numerical simulations performed during the past twenty years verify
that {\it major} mergers of {\it gas-rich} disk galaxies can plausibly
account for these phenomena and elucidate the underlying physics.
In \citet{hopkins:groups.qso}, 
we provide an outline of the phases of 
evolution that might be associated with a major merger in the 
lifetime of a massive galaxy, but we briefly summarize them here.
Tidal torques excited during a merger lead to rapid inflows of gas
into the centers of galaxies \citep{hernquist.89,barnes.hernquist.91,
barneshernquist96}, triggering starbursts \citep{mihos:starbursts.94,
mihos:starbursts.96} and feeding rapid black hole growth \citep{dimatteo:msigma}.
Gas consumption by the starburst and dispersal of residual
gas by supernova-driven winds and feedback from black hole growth 
\citep{springel:red.galaxies} terminate star formation so that the remnant
quickly evolves from a blue to a red galaxy.  The stellar component of
the progenitors provides the bulk of the material for producing the
remnant spheroid \citep{barnes:disk.halo.mergers,barnes:disk.disk.mergers,
hernquist:bulgeless.mergers,hernquist:bulge.mergers}
through violent relaxation. A major
merger is generally required in order for the tidal forces to excite a
sufficiently strong response to set up nuclear inflows of gas and build massive spheroids.
Although simulations suggest that the precise
definition of a major merger in this context is somewhat blurred by the
degeneracy between the mass ratio of the progenitors and the orbit of
the interaction \citep{hernquist.89,hernquist.mihos:minor.mergers,bournaud:minor.mergers},
systematic studies with both numerical simulations \citep{younger:minor.mergers} 
and observations \citep{dasyra:mass.ratio.conditions,woods:tidal.triggering} 
find that strong gas inflows and morphological transformation are typically only observed 
below mass ratios $\sim 3:1$, despite the greater frequency of 
higher mass-ratio mergers. In what follows, unless explicitly noted, we generally 
mean the term ``mergers'' to refer specifically to major mergers.

It also must be emphasized that essentially all numerical studies 
of spheroid kinematics find that {\em only} mergers 
can reproduce the observed kinematic properties of observed elliptical 
galaxies and ``classical'' bulges \citep{hernquist.89,hernquist:bulgeless.mergers,
hernquist:bulge.mergers,barnes:disk.halo.mergers,barnes:disk.disk.mergers,
schweizer92,naab:minor.mergers,bournaud:minor.mergers,
naab:gas,naab:dry.mergers,naab:profiles,jesseit:kinematics,cox:kinematics}. 
Disk instabilities and
secular evolution (e.g.\ bar instabilities, harassment, and other 
isolated modes) can indeed produce bulges, but these are invariably 
``pseudobulges'' \citep{schwarz:disk-bar,athanassoula:bar.orbits,
pfenniger:bar.dynamics,combes:pseudobulges,
raha:bar.instabilities,kuijken:pseudobulges.obs,oniell:bar.obs,athanassoula:peanuts}, 
with clearly distinct shapes (e.g.\ flattened or 
``peanut''-shaped isophotes), rotation properties (large $v/\sigma$), 
internal correlations (obeying different Kormendy and Faber-Jackson relations), 
light profiles (nearly exponential Sersic profiles), and colors and/or 
substructure from classical bulges 
\citep[for a review, see][]{kormendy.kennicutt:pseudobulge.review}. 
Observations indicate that 
pseudobulges constitute only a small fraction of the total mass density 
in spheroids \citep[$\lesssim10\%$; see][]{allen:bulge-disk,ball:bivariate.lfs,
driver:bulge.mfs}, becoming a large fraction of the bulge 
population only in small bulges in late-type hosts 
\citep[e.g.\ Sb/c, corresponding to typical $\mgal\lesssim10^{10}\,\msun$; see][and 
references therein]{carollo98, kormendy.kennicutt:pseudobulge.review}. 
This is not to say that secular processes cannot, in principle, 
build some massive bulges \citep[see e.g.][]{debattista:pseudobulges.a,debattista:pseudobulges.b}. 
However, although 
such processes may be important 
for the buildup of low mass black hole and spheroid 
populations, it is empirically clear that secular evolution {\em cannot} be the agent 
responsible for the formation of most 
elliptical galaxies. 

Motivated by these considerations, \citet{hopkins:qso.all,hopkins:red.galaxies} developed a
model where starbursts, quasars, supermassive black hole growth, and
the formation of red, elliptical galaxies are connected through an
evolutionary sequence, caused by {\it mergers} between {\it gas-rich}
galaxies. It is important to keep in mind
that this does not rule out other processes occurring at lower levels
and under other circumstances.  For example, we are not claiming that
all bulges result from mergers -- secular pseudobulge growth does 
appear to be important for small bulges in disk-dominated systems, 
and additional processes may act to redden satellite galaxies 
in massive halos, a potentially important contributor to the population 
of red galaxies at low masses $\mgal\lesssim10^{10}\,\msun$ \citep[e.g.][]{blanton:env}. 
Moreover, spheroid evolution by gas-free (``dry'') mergers will go on, but does
not explain how stellar mass is initially moved onto the red sequence 
or transformed from disk to spheroid. 

All of this, however, only goes to the question of the formation of elliptical 
galaxies, not to the question of how such galaxies become (and stay) 
``red and dead.'' It is well established from both numerical simulations 
\citep{springel:red.galaxies} and observations \citep[e.g.][]{rothberg.joseph:kinematics}
that merger remnants redden rapidly onto the 
red sequence as typical early-type galaxies. 
However, it is still debated whether or not such systems 
will stay on the red sequence for long periods of time, since this requires some 
suppression of subsequent accretion and cooling as their host 
dark matter halos grow. In massive elliptical galaxies, it is not obvious how 
the formation of cooling flows has been suppressed since $z\sim2$, 
despite observations finding that the cooling times of large quantities of 
gas are shorter than a Hubble time. 
In other words, there is an important outstanding question, 
which we seek to address: do major mergers 
or their remnants effectively quench future star formation (i.e.\ 
maintain low star formation rates for significant cosmic times), or is 
it some other, independent process which is responsible for quenching? 

At low redshift, there appears to be a clear association between quenched 
(red, passive) galaxies and the presence of a massive spheroid, at least 
for the relatively massive $\mgal\gtrsim10^{10}\,\msun$ systems of interest in 
this paper. \citet{bell:mfs} and \citet{mcintosh:size.evolution} 
find that $\gtrsim80\%$ of the $z=0$ red population are classical, bulge-dominated 
systems, with most of the remainder being early-type disks. 
\citet{drory:quenching.vs.bulge.type} further 
investigate these disk-dominated systems, and find that early-type disks on the 
red sequence have uniformly classical bulges (presumably formed via mergers), 
whereas disks of comparable mass, luminosity, and bulge size hosting 
pseudobulges (formed via secular instabilities) remain in the blue cloud. At 
higher redshifts, morphological signatures are less clear, and an increasingly large fraction 
of red galaxies (naively identified by simple color cuts) are contaminant dusty or 
edge-on disks (clearly not true quenched/passive systems). However, those 
systems which can be clearly identified as truly 
passive appear to be overwhelmingly compact spheroids \citep{mcintosh:size.evolution,bundy:mfs}, 
even at $z\sim2-3$ \citep{labbe05:drgs,kriek:drg.seds,zirm:drg.sizes}. 
This suggests a strong connection between 
a major, spheroid-forming merger and galaxy quenching.

The standard framework for understanding quenching follows the 
cooling of gas in the galaxy host halo. From simple 
scaling arguments one can show that at low halo masses 
the cooling time will (in the absence of heating 
mechanisms) be shorter than the free-fall time of the gas, and 
accretion is only limited by the free-fall of newly accreted halo 
gas onto the central galaxy -- the 
so-called ``rapid cooling'' or ``cold accretion'' regime. Once the 
halo becomes sufficiently massive, the cooling time becomes 
longer than the free fall time, and so gas does not simply fall 
onto the central galaxy, but rather forms a quasi-static, pressure 
supported hydrostatic equilibrium -- the ``hot halo'' regime. New gas accreted will shock 
against this pressure-supported structure, heating itself and the 
gas interior to it, and 
accretion will proceed only gradually, from the cooling of the gas at the 
center of the halo \citep{rees.ostriker.77,
norman.silk:gas.halos,blumenthal.84}. 

Numerical simulations suggest that this transition 
occurs at a mass $\mhalo\sim10^{11}-10^{12}\,\msun$ \citep{birnboim:mquench,keres:hot.halos}. 
In many prescriptions (such as the ``halo quenching'' models to which we refer in 
\S~\ref{sec:ellipticals:fractions}), it is simply assumed that the development of a hot halo 
at this mass threshold is the dominant criterion for quenching. However, 
both numerical simulations and analytic calculations 
\citep[][and references therein]{kh00,benson:sam,keres:hot.halos} argue that 
this transition alone cannot solve the ``cooling flow'' problem -- namely that the 
high densities at the core of the pressure-supported hot halo will allow 
rapid cooling onto the central galaxy, producing large galaxies which 
are much too massive, gas-rich, disk-dominated, actively star-forming, young and blue 
relative to the observations. Some kind of 
heating term is needed to prevent this from occurring and 
maintain quenching. 

It has become popular to invoke activity from a 
low-Eddington ratio AGN in the central galaxy as the source of this heating term -- 
the ``radio-mode'' AGN \citep{croton:sam}. This, however, 
requires the presence of a massive black hole (and therefore 
a correspondingly massive spheroid) accreting in a relatively 
low steady state (i.e.\ with most of the cold gas in the galaxy consumed). This requirement, 
along with the arguments above, suggests that merger history might be just as important 
as (if not more important than) halo mass in determining the quenching of a given galaxy. 
Ultimately, we emphasize that the detailed numerical simulations and 
analytic calculations of the hot halo regime do {\em not} argue that 
entering this regime does, or can, directly quench future cooling. Rather, these 
calculations argue only that the ``hot halo'' regime provides an ideal environment 
{\em in which quenching mechanisms might operate}. 

Unfortunately, obtaining a purely theoretical framework for 
any quenching scenario is difficult
because cosmological simulations including gas dynamics currently lack
the resolution to describe the small-scale physics associated with
disk formation, galaxy mergers, star formation, and black hole growth. 
A popular alternative has been the employment of semi-analytic methods, 
adopting various prescriptions for quenching and feedback 
processes and comparing the predictions with observed 
galaxy populations \citep[e.g.][]{kh00,somerville:sam,benson:sam,khochfar:sam,
granato:sam,scannapieco:sam,kang:sam,delucia:sam,monaco:sam}
These models have robustly shown
the need for some quenching processes, and their great success has been 
demonstrating that simple prescriptions for basic feedback elements 
yield good agreement with local galaxy mass/luminosity functions and 
color distributions \citep[e.g.][]{croton:sam,bower:sam,cattaneo:sam}.

However, the similar success of a large variety of 
such prescriptions at matching these basic local constraints has 
demonstrated that such predictions are fundamentally
{\em non-unique}. For example, \citet{cattaneo:sam} have shown that 
one obtains similar galaxy mass functions and color-magnitude 
relations whether one adopts a pure halo mass threshold 
for quenching, a halo mass threshold which depends on 
some feedback balance with a low-luminosity AGN, 
or a (halo mass-independent) 
galaxy bulge-to-disk criterion. Clearly, these simple constraints are 
insufficient to discriminate between the mechanisms associated 
with galaxy quenching. Furthermore, the diversity of semi-analytic 
prescriptions has demonstrated that there are considerable 
degeneracies between, for example, the prescriptions for star formation in disks 
and those for quenching, despite the fact that the two should be 
constrained by independent galaxy populations. It is therefore 
necessary to determine what, if any, are the robust differences between 
various quenching prescriptions, and to study higher-order 
observational constraints (such as e.g.\ the redshift evolution of 
populations, or bivariate distributions of galaxy properties as a function of 
both galaxy mass and halo mass or galaxy kinematics) that 
hold the potential to break these degeneracies. 

In the first of a pair of companion papers 
\citep[][henceforth \paperone]{hopkins:groups.qso}, 
we describe a strategy that enables us, for the first
time, to provide a purely theoretical framework for our models of 
merger-induced activity. 
By combining previous estimates of the evolution of the halo mass
function with halo occupation models and our estimates for merger
timescales, we infer the statistics of mergers that 
form spheroids. Because our merger simulations relate starbursts,
quasars, and red galaxies as different phases of the same events, we
can graft these simulations onto our theoretical, cosmological 
calculation and determine the cosmological birthrate of these various
populations and their evolution with redshift.  In particular we
demonstrate in what follows that there are a number of unique, 
robust predictions of a model in which mergers drive the quenching 
of galaxies (in addition to forming spheroids in the first place), 
distinct from the predictions of models in which this quenching is set just by 
halo properties or secular (disk) instabilities. We find that observations 
of red galaxies support our predictions, and disfavor 
other theoretical models. 

In \paperone, 
we describe our model and use it to investigate the properties of 
mergers and merger-driven quasar activity. 
In this paper (\papertwo), 
we extend this to study the properties of merger remnants and the 
formation of the early-type galaxy population.
We begin by briefly reviewing the key 
elements of the model from \paperone\ in 
\S~\ref{sec:mergers}. 
In \S~\ref{sec:ellipticals} we use the method developed in 
\paperone\ to examine the 
consequences of a general model in which 
major merger remnants remain ``quenched'' once the merger 
terminates star formation. Specifically, 
\S~\ref{sec:ellipticals:int} shows the predictions of this 
model for the buildup of early-type or red galaxy mass functions 
and mass density with redshift, and the formation times of 
early-type galaxies. In \S~\ref{sec:ellipticals:fractions} we demonstrate 
how the resulting fraction of red or ``quenched'' galaxies depends on 
properties such as halo and galaxy mass, and contrast these 
with the predictions of alternative models in which the 
quenching is associated with a halo mass criterion or 
secular processes (disk instabilities). In \S~\ref{sec:ellipticals:evolution} 
we extend these comparisons to the redshift evolution of these trends.
In \S~\ref{sec:ellipticals:dry} we
briefly examine the role of subsequent gas-poor 
major mergers in this model, and compare with observations of 
early-type galaxy structure. 
In \S~\ref{sec:quenching} we outline the broad physical 
mechanisms which give rise to such a model. We examine 
in \S~\ref{sec:quenching:transition} 
how mergers are associated with the ``transition'' of 
galaxies from the blue cloud to the red sequence, and 
in \S~\ref{sec:quenching:maintenance} we examine the 
role of different feedback mechanisms in ``maintaining'' low 
star formation rates in remnant elliptical galaxies. 
We discuss and summarize our 
conclusions in \S~\ref{sec:discussion}. 

Throughout, we adopt a WMAP3 
$(\Omega_{\rm M},\,\Omega_{\Lambda},\,h,\,\sigma_{8},\,n_{s})
=(0.268,\,0.732,\,0.704,\,0.776,\,0.947)$ cosmology 
\citep{spergel:wmap3}, and normalize all observations and models 
shown to this cosmology. 
Although the exact choice of 
cosmology may systematically 
shift the inferred bias and halo masses (primarily scaling with $\sigma_{8}$), 
our comparisons (i.e.\ relative biases) are for the most part unchanged, 
and repeating our calculations for 
a ``concordance'' $(0.3,\,0.7,\,0.7,\,0.9,\,1.0)$ cosmology or 
the WMAP1 $(0.27,\,0.73,\,0.71,\,0.84,\,0.96)$ results of \citet{spergel:wmap1}
has little effect on our conclusions. 
We also adopt a diet Salpeter IMF following \citet{bell:mfs}, and convert all stellar masses 
and mass-to-light ratios to this choice. Again, the exact choice of IMF systematically 
shifts the normalization of stellar masses herein, but does not substantially change 
our comparisons. 
$UBV$ magnitudes are in the Vega system, and 
SDSS $ugriz$ magnitudes are AB.

\section{Mergers: The Basic Model}
\label{sec:mergers}

The model which we use to calculate the rate and nature of mergers 
as a function of e.g.\ mass, redshift, and environment is described in 
detail in \paperone, but we briefly outline the key elements here. 

{\bf 1.\ Halo Mass Function:} We begin by adopting the halo mass function 
following \citet{shethtormen}. 
There is little ambiguity in this calculation at all redshifts and masses 
of interest \citep[$z\lesssim6$; e.g.][]{reed:halo.mfs}, and we do not consider it a significant source of 
uncertainty. 

{\bf 2.\ Subhalo Mass Function:} The subhalo mass function of each halo is 
then calculated. Although numerical simulations and semi-analytic 
calculations generally give 
similar results \citep[especially for the major-merger mass ratios of interest 
in this paper, as opposed to very small subhalo populations; see][]{vandenbosch:subhalo.mf}, 
there is still some (typical factor $<2$) disagreement between different estimates. 
We therefore repeat most of our calculations adopting both 
our ``default'' subhalo mass function calculation 
\citep{zentner:substructure.sam.hod,kravtsov:subhalo.mfs} and an alternative 
subhalo mass function calculation \citep{vandenbosch:subhalo.mf} 
\citep[normalized to match cosmological simulations 
as in][]{shaw:cluster.subhalo.statistics}, which bracket the range 
of a number of different estimates \citep[e.g.,][]{springel:cluster.subhalos,
tormen:cluster.subhalos,delucia:subhalos,gao:subhalo.mf,nurmi:subhalo.mf} 
and demonstrate the uncertainty 
owing to this choice. The difference is ultimately negligible 
at $\mgal\gtrsim10^{10}\,\msun$ (where, unless otherwise specified, $\mgal$ 
refers to the baryonic mass of the galaxy)
at all redshifts, and rises to only a factor $\sim2$ at 
$\mgal\lesssim10^{10}\,\msun$ (probably owing to differences in the 
numerical resolution of various estimates at low halo masses). 

{\bf 3.\ Halo Occupation Model:} We then populate the 
central galaxies and ``major'' subhalos with an empirical halo occupation model. 
Although such models are constrained, by definition, to reproduce the mean 
properties of the halos occupied by galaxies of a given mass/luminosity, there 
are known degeneracies between parameterizations that give rise to 
(typical factor $\sim2$) differences between models. We therefore again 
repeat all our calculations for our ``default'' model 
\citep{conroy:monotonic.hod} \citep[see also][]{valeostriker:monotonic.hod} and 
an alternate halo occupation model \citep{yang:clf} \citep[see also][]{yan:clf.evolution,zheng:hod}, which 
bracket the range of a number of calculations \citep[e.g.,][]{vandenbosch:concordance.hod,
cooray:highz,cooray:hod.clf,zheng:hod}. Again, we find this
yields negligible differences 
at $\mgal\gtrsim10^{10}\,\msun$ (as the clustering and abundances 
of massive galaxies are reasonably well-constrained, and most of these 
galaxies are central halo galaxies), and even at low masses the 
typical discrepancy in our predictions owing to the 
choice of halo occupation model rises to only $\sim0.2\,$dex. 

We note that we have also considered a variety of prescriptions for the 
redshift evolution of the halo occupation model: including that 
directly prescribed by the quoted models, a complete re-derivation 
of the HOD models of \citet{conroy:monotonic.hod} and 
\citet{valeostriker:monotonic.hod} 
at different redshifts from the observed mass functions of 
\citet{bundy:mfs,fontana:highz.mfs,borch:mfs,blanton:lfs} (see \paperone), 
or assuming no evolution (in terms of galaxy mass
distributions at fixed halo mass; for either all galaxies or 
star-forming galaxies). We find that the resulting differences are 
small (at least at $z\lesssim3$), comparable to 
those inherent in the choice of halo occupation model. 
This is not surprising, as a number of recent 
studies suggest that there is little evolution in halo occupation 
parameters (in terms of mass, or relative to $L_{\ast}$) with 
redshift \citep{yan:clf.evolution,cooray:highz,
conroy:monotonic.hod}, or equivalently that the masses of galaxies hosted in a 
halo of a given mass are primarily a function of that halo mass, not 
of redshift \citep{heymans:mhalo-mgal.evol,
conroy:mhalo-mgal.evol}. This appears to be especially true for 
star-forming and $\sim L_{\ast}$ galaxies \citep[of greatest importance for 
our conclusions;][]{conroy:mhalo-mgal.evol}, unsurprising 
given that quenching is not strongly operating in those systems to change 
their mass-to-light ratios. 

{\bf 4.\ Merger Timescale:} Having populated a given halo and its subhalos 
with galaxies, we then calculate the timescale for mergers between major galaxy 
pairs. This is ultimately the largest source of uncertainty in our calculations, 
at all redshifts and masses. 
Again, we emphasize that some of our calculations are completely 
independent of these timescales. However, where adopted, we illustrate  
this uncertainty by presenting all of our predictions for three estimates of 
the merger timescale: 
first, a simple dynamical friction formula (this is what is generally 
adopted in semi-analytic models, for example). Second, a 
group capture or collisional (i.e.\ effective 
gravitational) cross section \citep[e.g.][]{white:cross.section,krivitsky.kontorovich,
makino:merger.cross.sections,mamon:groups.review} 
approximation, generally more appropriate on small scales, 
in satellite-satellite mergers, or in the merger 
of two small field halos. Third, an angular 
momentum (orbital cross section) capture estimate \citep[i.e.\ 
considering capture into the effective angular-momentum space 
of mergers;][]{binneytremaine}. 

At large masses 
and redshifts $z\lesssim2.5$, this is a surprisingly weak source of 
uncertainty, but the estimated merger rates/timescales 
can be different at low masses $\mgal\lesssim 10^{10}\,\msun$ 
and the highest redshifts $z\sim3-6$. At low masses, this owes 
to a variety of effects, including the substantial difference 
between infall or merger timescales and the timescale for 
morphological disturbances to be excited (different in e.g.\ an 
impact approximation as opposed to the circular orbit decay 
assumed by dynamical friction). Note that where relevant, we have used  
numerical simulations to estimate the typical duration of the final merger 
stages or e.g.\ the morphological relaxation time  
\citep[in which mergers will be identified by typical 
morphological classification schemes, see][]{lotz:merger.selection}.
The difference in redshift 
evolution is easily understood: at fixed mass ratio, the 
dynamical friction timescale scales as 
$t_{\rm df}\propto \tH\propto \rho^{-1/2}$, 
but a ``capture'' timescale will scale with fixed cross section as 
$t\propto 1/(n\,\langle\sigma\,v \rangle)\propto \rho^{-1}$, 
so that (while the details of the cross-sections make the 
difference not quite as extreme as this simple scaling) the very high densities at 
high redshift make collisional merging grow rapidly in efficiency. 
The true solution is probably some effective 
combination of these two estimates, and the 
``more appropriate'' approximation 
depends largely on the initial orbital parameters of the subhalos. 
At present, we therefore must recognize this as an inherent 
uncertainty, but one that serves to bracket the likely range of 
possibilities at high redshifts. 

In \paperone\ (\S~2.2), we show that 
together, these criteria naturally define a preferred major-merger scale (host halo mass $\mhalo$) for 
galaxies of mass $\mgal$ -- the ``small group scale,'' only slightly larger than 
the average halo hosting a galaxy of mass $\mgal$. This is the scale at which 
the probability to accrete a second galaxy of comparable mass $\sim\mgal$ (fuel for a 
major merger) first becomes significant. At smaller (relative) 
halo masses, the probability that the halo 
hosts a galaxy as large as $\mgal$ declines rapidly. At larger masses, the 
probability that the halo will merge with or accrete another halo hosting a comparable $\sim\mgal$ 
galaxy increases, but the efficiency of the merger of these galaxies declines rapidly. 
We stress that this small group scale is distinct from the more typical large group scale 
identified observationally (the average small group halo will still host only 1 galaxy 
of mass $\sim\mgal$, and groups will only consist of $2-3$ members of similar mass). 
This is not to say, however, that mergers occur (in a global sense) at a specific scale, 
since the small group scale is different for different galaxy masses -- 
a consequence of this model is the observational fact that mergers occur in halos of 
all masses and in all environments
\citep[including field and even void environments;][]{alonso:groups,
goto:e+a.merger.connection,hogg:e+a.env}, although 
the characteristic masses 
and star formation histories 
of galaxies merging will change in different environments. 

In \paperone\ we compare this model with a number of observations, and 
show that it reproduces the mass functions and star formation 
histories of galaxies, merger mass functions (and infrared 
luminosity functions) and merger fractions as 
a function of galaxy and/or halo mass and redshift, the clustering of 
mergers as a function of mass and redshift, and the dependence of 
merger rates and fractions on small-scale environmental properties. 
This provides some reassurance that we are accurately predicting 
the rate and nature of major mergers as a function of 
these properties, and can use this model to make robust predictions 
for the nature of merger remnants.

\section{Ellipticals}
\label{sec:ellipticals}

We now turn to the possibility of an 
association between mergers and the termination or quenching of 
star formation in remnant galaxies. 
In \S~\ref{sec:quenching} we consider 
potential physical mechanisms for this quenching, but we 
caution that at present these mechanisms are neither well-understood 
nor observationally well-constrained. As a consequence, we first wish to 
examine the consequences of the simple hypothesis that {\em some} 
mechanism quenches star formation after a major merger, whether
it involves gas exhaustion, starburst or quasar feedback, hot halo formation, 
or other mechanisms. We therefore 
make the simple ansatz: {\em Systems are 
quenched after a major merger of star-forming/gas-rich galaxies}. 

\subsection{Integrated Populations}
\label{sec:ellipticals:int}

\begin{figure}
    \centering
    \figexpand
    \plotone{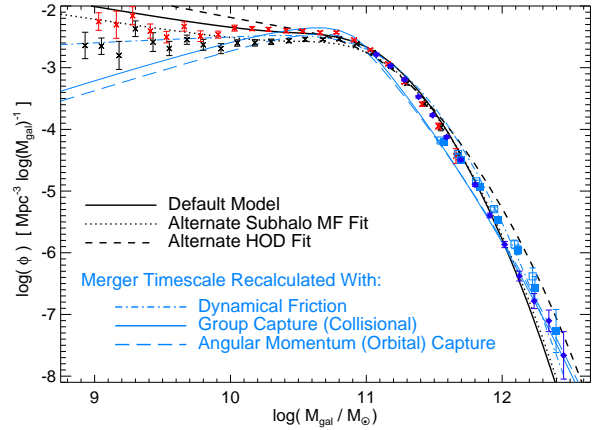}
    \caption{Predicted local quenched/red/early-type galaxy 
    mass function (lines) obtained by integrating 
    forward the major merger mass function to $z=0$ (i.e.\ assuming 
    that each merger leaves a quenched early-type remnant). 
    Different styles show different 
    variants of our calculation which bracket the 
    range of our uncertainties, varying e.g.\ the subhalo mass functions, 
    halo occupation model, and approximation used to calculate merger 
    timescales (as described in \S~\ref{sec:mergers}; see \paperone\ for 
    a more detailed comparison). 
    We compare with observed early-type or red galaxy mass functions 
    from \citet[][SDSS elliptical and red galaxy mass functions; 
    black and red $\times$'s, respectively]{bell:mfs}, 
    \citet[][SDSS LRGs; blue squares]{wake:lrgs}, and 
    \citet[][6dF LRGs; purple diamonds]{jones:lrgs}.
    The mass functions from \citet{wake:lrgs} and \citet{jones:lrgs} are 
    converted from luminosity functions using the luminosity-dependent 
    mass-to-light ratios from \citet{bell:mfs}. We show both the 
    directly measured \citet{wake:lrgs} result (open) and that corrected for 
    passive evolution from $z=0.1$ (filled). 
    \label{fig:local.mf}}
\end{figure}
\begin{figure}
    \centering
    \figexpand
    \plotone{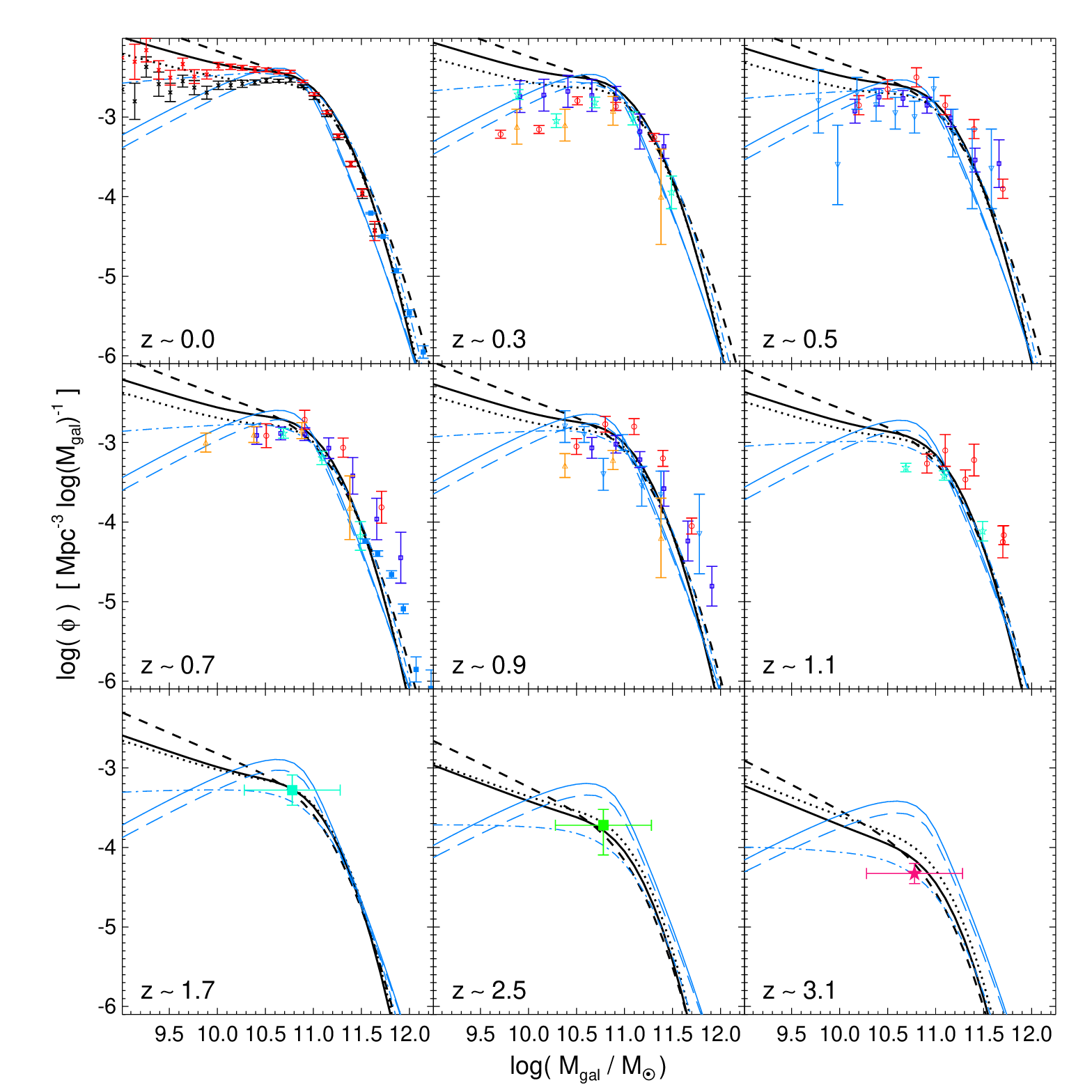}
    \caption{As Figure~\ref{fig:local.mf}, but at each of several redshifts. Points 
    at $z=0$ are as in Figure~\ref{fig:local.mf}. At higher redshifts, early-type or 
    red galaxy MFs are shown from \citet[][red circles]{bundy:mfs,bundy:mtrans}, 
    \citet[][purple squares]{borch:mfs}, \citet[][cyan stars]{franceschini:mfs}, 
    \citet[][orange triangles]{pannella:mfs}, \citet[][blue inverted triangles]{fontana:mfs}, 
    \citet[][blue filled squares]{wake:lrgs}, and points 
    at $z=1.7,\,2.5$, and $3.1$ are estimated from the number density of passively evolving 
    (non-star forming) red galaxies with stellar masses $\gtrsim10^{11}\,\msun$ 
    in \citet[][cyan square]{daddi05:drgs}, \citet[][green square]{labbe05:drgs}, and 
    \citet[][magenta star]{grazian:drg.comparisons}, 
    respectively. Masses (or mass ranges) have been corrected to our adopted IMF. The 
    integrated merger mass function is consistent with the observed red galaxy mass 
    function at all redshifts $z\sim0-3.5$. 
    \label{fig:redshift.mf}}
\end{figure}

In \paperone\ we calculated the 
major merger rate of galaxies as a function of 
galaxy mass and redshift. If each such merger leaves a quenched early-type 
remnant, then
we can integrate the merger rate forward in time to 
obtain the early-type or red galaxy mass function at each redshift, 
\begin{equation}
\label{eqn:mf.int}
\phi_{\rm early}(\mgal) = \int{\dot{n}(\mgal\,|\,z)}\,\frac{{\rm d}t}{{\rm d}z}\,{\rm d}z.
\end{equation}
Figures~\ref{fig:local.mf} \& \ref{fig:redshift.mf} show this at several redshifts
for our model of major mergers. 
Note that Equation~\ref{eqn:mf.int} adds the contribution from all mergers -- 
i.e.\ implicitly includes in the mass function the contribution from ``dry'' or 
spheroid-spheroid mergers. Technically, we should also include the 
sink term from dry mergers, 
$-2\,\int{\dot{n}_{\rm dry}(0.5\,\mgal\,|\,z)\,{\rm d}t}$, representing the loss of 
two early-types of mass $\sim\mgal/2$ for each major dry merger of final mass $\mgal$. 
This requires a number of additional assumptions for the red/blue galaxy fraction 
as a function of $\mgal$ or $\mhalo$ and the 
initial mass ratios of mergers, so we have not included it here, 
but note that for reasonable empirical 
estimates (such as those in \S~\ref{sec:ellipticals:fractions}) of these numbers, the sink 
term has little effect. That is not to say that at low redshift, 
the dry merger contribution cannot indeed be important to the shape of the mass function
where it is falling steeply at high mass ($\gg\mstar$). Because of this 
steep fall-off, moving a small fraction of lower mass systems to higher masses 
can significantly increase the number density of the most massive systems. 
However, the loss of 
less massive systems is a small correction. The dominant term 
at masses ($\lesssim$ a few $\mstar$) important for the total mass density 
of red systems is the movement of systems 
to the red sequence by gas-rich mergers. 

We have also neglected 
growth via minor mergers: however, we demonstrate in \paperone\ 
that this is also a small correction; i.e.\ mass growth is 
dominated by major mergers and star formation, as seen in 
cosmological simulations \citep{maller:sph.merger.rates} and 
observations \citep{zheng:hod.evolution} (although it is possible that minor mergers 
become important for the most extreme, massive BCGs). 

As discussed in \S~\ref{sec:mergers}, there are a number of uncertainties at 
the lowest masses $\mgal\lesssim10^{10}\,\msun$, which are evident 
in the differences between our predictions 
in Figures~\ref{fig:local.mf} \&\ \ref{fig:redshift.mf} -- these include issues of 
completeness and resolution in the subhalo mass functions and 
halo occupation models, and sensitivity (for very low-mass mergers) to the 
method used to calculate merger cross sections (for example, 
the difference between a dynamical friction and an impact approximation becomes 
large). The predictions in this regime are probably subject to a number of 
other caveats, as well. At the lowest masses $\mgal\lesssim{\rm a\ few}\times10^{9}\,\msun$, 
satellite-satellite mergers (the dynamics of which are sensitive to orbital parameters) 
become an important contributor to the 
total merger remnant population. Also, the fraction of observed pseudobulges 
starts to become large, implying that secular instabilities may begin contributing 
significantly to the early-type population below these masses. Finally, many 
of the observed red galaxies at masses below this threshold (almost an 
order of magnitude below $\mstar$) are satellites of more massive systems, so 
processes like ram pressure stripping, tidal stripping, harassment, and a 
cutoff of new accretion are likely to be important (and may even dominate 
their becoming red in the first place). 
At higher masses $\mgal\gtrsim10^{10}\,\msun$, however, 
the agreement 
between our predictions is good, 
regardless of which subhalo mass functions, halo occupation models, 
or merger timescale approximations we adopt. Moreover,
almost all of these galaxies are observed to be central halo galaxies 
\citep[e.g.][]{weinmann:obs.hod} and the pseudobulge fraction 
is small, so we can have some confidence 
that satellite and secular processes are not a large effect 
(see also \S~\ref{sec:ellipticals:fractions}). 
We refer to \paperone\ for a 
more detailed comparison, but we have tested the model extensively for 
these masses ($\mgal\gtrsim10^{10}\,\msun$), 
and find it is both robust and consistent with observed statistics of 
mergers as a function of galaxy and halo mass, redshift, and 
galaxy color/morphological type. 

\begin{figure}
    \centering
    \figexpand
    \plotone{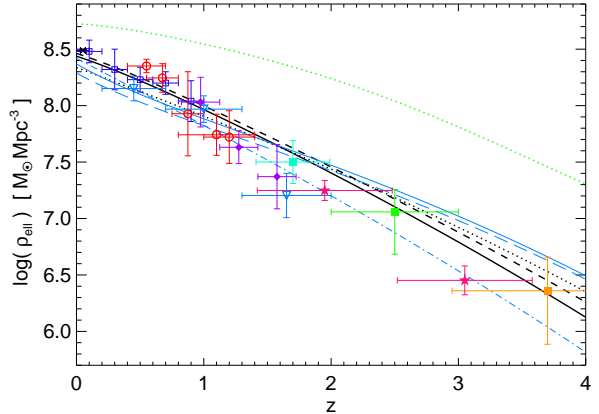}
    \caption{Integrated stellar mass density of red or early-type 
    (passive) galaxies as a function of redshift, 
    compared to the predicted stellar mass density which has undergone a major galaxy merger 
    (i.e.\ integrating the merger rate function of star-forming galaxies). 
    Points are as in Figures~\ref{fig:local.mf} \& \ref{fig:redshift.mf}, with 
    lines showing the predictions from our 
    different methods of calculating the merger rate function from \S~\ref{sec:mergers} (see \paperone). 
    We add the mass density estimates from 
    \citet[][violet diamonds]{abraham:red.mass.density} and \citet[][orange square]{vandokkum06:drgs}.
    Dotted green (uppermost) line shows the total stellar mass density of the universe expected 
    from the integrated star formation history in 
    \citet{hopkinsbeacom:sfh} (normalized to the $z=0$ value from \citet{bell:mfs}). 
    The mass density of systems which have undergone major, gas-rich mergers 
    agrees well at all redshifts with the mass density in red or early-type galaxies, with 
    sufficient mergers occurring at high redshifts to account for the observed densities 
    from \citet{labbe05:drgs} and \citet{grazian:drg.comparisons} at $z\sim2-4$. 
    \label{fig:red.mass.density}}
\end{figure}

Figure~\ref{fig:red.mass.density} plots the integrated version of this, namely the 
mass density of red/early-type systems as a function of redshift. Here, we 
integrate only the gas-rich merger rate function (the same merger rate function we 
used to predict the quasar luminosity function in \paperone; i.e.\ using our 
empirical halo occupation model to identify specifically mergers of gas-rich or 
star-forming galaxies), as 
dry mergers cannot, by definition, increase the mass density of red galaxies. 

The integrated 
mass which has undergone major, gas-rich mergers agrees well with the mass 
density of red galaxies at all redshifts. Even at high redshifts 
$z\sim2-4$, this merger-driven model has no difficulty accounting for the 
relatively large mass densities of red galaxies observed by e.g.\ \citet{labbe05:drgs}, 
\citet{vandokkum06:drgs}, \citet{grazian:drg.comparisons}, and \citet{kriek:drg.seds}, 
as the highest-overdensity peaks in the early universe undergo rapid 
major mergers \citep[suggested in the observations as well, given 
the color-density relation of these objects;][]{quadri:highz.color.density}.
It is important to note that a significant number of these high-redshift systems 
have been spectroscopically confirmed as passive, ``red and dead'' systems 
\citep{kriek:drg.seds,wuyts:irac.drg.colors} 
with elliptical morphologies \citep{zirm:drg.sizes} and relatively 
old ages ($\sim{\rm a\ few}\times10^{8}\,$yr, or $\sim1/5\,\tH$ at these redshifts). 
This is in strong contrast to some pure hot-halo quenching models, in 
which cold accretion within a hot halo persists at high redshifts $z\gtrsim2$. 

We consider a detailed comparison with these models in \S~\ref{sec:ellipticals:evolution}, 
but note, for example, 
that the \citet{dekelbirnboim:mquench} estimate of the hot halo quenching mass 
predicts that at $z\sim3.5$, cold flows continue within all $\mhalo \lesssim10^{14}\,\msun$ 
halos, which allows only a completely negligible maximum 
red galaxy mass density (even if we adopt a $100\%$ baryon 
conversion efficiency and assume all quenched halos are ``red'') -- even lowering 
this threshold by an order of magnitude predicts a quenched galaxy mass density 
an order of magnitude below that observed.

\begin{figure}
    \centering
    \figexpand
    \plotone{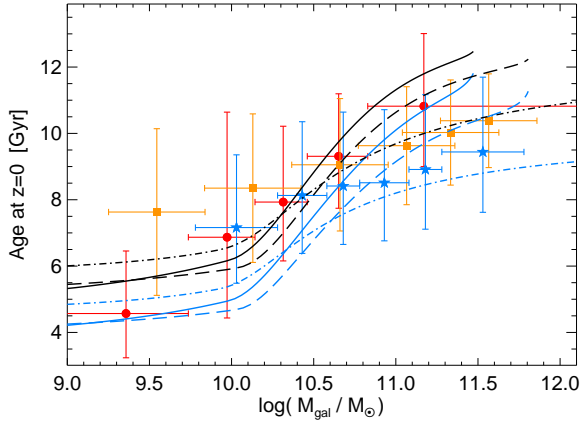}
    \caption{Predicted ages of early-type galaxies (at $z=0$) as a function of 
    stellar mass from the integrated mass functions in Figure~\ref{fig:redshift.mf}, 
    compared to observations from \citet[][red circles]{nelan05:ages}, 
    \citet[][orange squares; we take their mean values as opposed to 
    those in a specific environment]{thomas05:ages}, and \citet[][blue stars]{gallazzi06:ages}. 
    Errors show the mass ranges and dispersion in ages within each mass range (not 
    the error in the mean ages). Blue lines show the predicted mean lookback time  
    to the final gas-rich merger for different 
    estimates of the merger rate as in Figure~\ref{fig:local.mf}. Black lines show 
    the same, but with age calculated from the mean redshift (as opposed to 
    lookback time), showing the systematic offset owing to this choice of definition. 
    \label{fig:ages}}
\end{figure}

Having integrated forward the implied rate of formation of early-type galaxies, we 
can also predict the ages of early types as a function of their mass. Figure~\ref{fig:ages} 
shows this comparison. We note that age here is defined as the time since 
the last gas-rich major merger (systems may, of course, undergo 
subsequent gas-free mergers, but this will not contribute new star formation). 
Our model contains no information 
about the prior star formation histories of merging disks. However, the observations 
to which we compare typically measure single stellar population (single burst) or 
light-weighted ages, which tend to reflect the last significant epoch of 
star formation. 

We emphasize that this does {\em not} imply that most of the stars in spheroids 
form in a short-lived, merger-induced burst. 
Direct calculation of the inferred stellar population 
ages from line index and SED fitting \citep[following][]{trager:ages} 
for realistic star formation histories 
from the semi-analytic models of \citet{somerville:sam} and the
hydrodynamical 
merger simulations of \citet{robertson:fp} suggests that the 
ages inferred for present early-type galaxies indeed reflect the epoch of 
the termination of star formation, even when $\gtrsim95\%$ of stars are 
formed over a much longer timescale at significantly earlier times 
(in these cases, in quiescent star formation in disks). 

Since we are interested in testing the possibility that major, gas-rich 
mergers are associated with the {\em termination} of star formation, these ages are the 
most appropriate with which to compare. But we again emphasize the caveat that 
without the details of the star formation histories in progenitor disks, our 
ages are subject to some systematic uncertainties. 
In any case, the agreement is good, 
suggesting that mergers have the correct timing, in a cosmic sense, to 
explain the shutdown of star formation in early-type systems.

\subsection{Color-Density Relations: The Dependence of Red Fractions on 
Halo Mass and Environment}
\label{sec:ellipticals:fractions}

\begin{figure*}
    \centering
    \figexpand
    \plotone{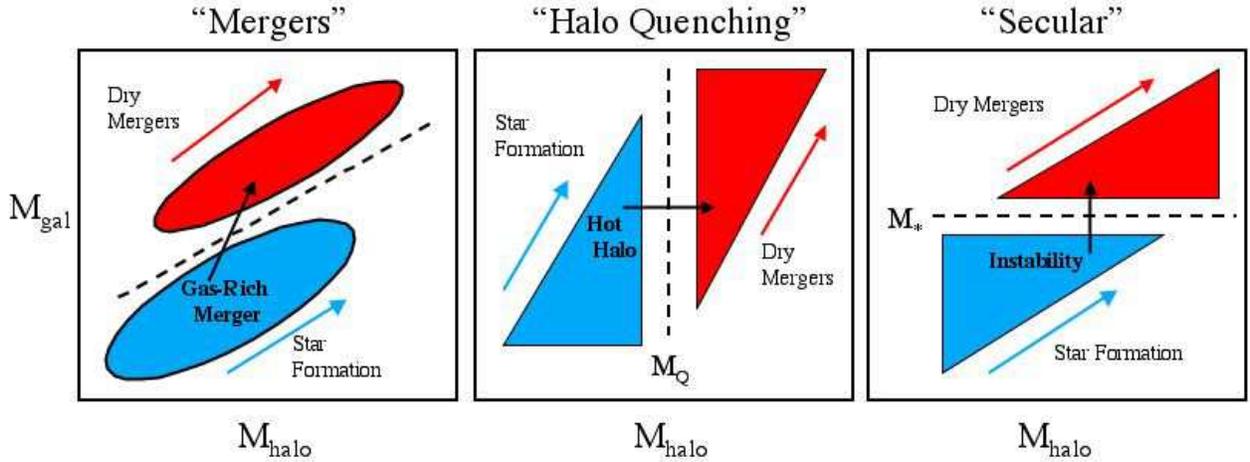}
    \caption{Qualitative illustration of galaxy growth and quenching in three different 
    basic models: a ``merger'' model, in which systems are quenched (for 
    any reason) after a major, gas-rich merger; a ``halo quenching'' model, in which 
    systems are uniformly quenched when their halo reaches a critical mass 
    $M_{Q}$ and establishes a ``hot halo'' gas accretion mode; and a ``secular'' model, 
    in which internal galactic processes (e.g.\ instabilities) determine 
    and color, independent of external processes. In all three models, star formation 
    and accretion move systems to larger galaxy and halo masses in the blue cloud 
    (blue shaded regions), and dry mergers move systems to larger masses 
    in the red sequence (red shaded regions). However, the division in this 
    galaxy-halo mass space is different in each case: for the ``halo quenching'' or 
    ``secular'' cases it depends solely on halo mass or galaxy mass, respectively. 
    In the ``mergers'' case, the transition line is tilted, as the probability of 
    mergers depends both on galaxy and halo mass. More massive halos 
    are more evolved, live in higher-density regions, and have more likely accreted 
    other galaxies to supply a major merger, so the red fraction increases with halo mass. 
    But at a given $\mhalo$, mergers are more efficient for high-mass systems 
    (and initial capture more likely), so the red fraction increases with galaxy mass. 
    Note that for all of these, we are explicitly focused on {\em central} galaxies, 
    and ignore processes that may redden satellites.
    \label{fig:cartoon}}
\end{figure*}

We now study the distribution of red galaxies in greater detail, to highlight the 
{\em unique} features of a merger-driven quenching model. For clarity, we focus 
only on {\em central} galaxies, and ignore the (potentially) completely 
physically distinct mechanisms (ram pressure stripping, tidal heating, etc.) 
responsible for quenching satellite galaxies in massive halos. 
Therefore, in what follows, our comparison of 
quenching and red/blue galaxies explicitly ignores satellite galaxies. 

Figure~\ref{fig:cartoon} 
illustrates three qualitatively distinct classes of models for quenching. 
We distinguish our ``merger'' model (systems quench after a major, gas-rich merger), 
a pure ``halo quenching'' model (systems quench upon crossing a critical 
halo mass), and a ``secular'' model (internal processes -- set by the galaxy mass 
and/or size -- 
solely determine galaxy color/star formation history). The models all predict that the 
most massive halos and most massive galaxies are predominantly quenched. 
However, in detail, the models differ in the behavior of quenching with 
respect to galaxy and halo mass.

In the simplest halo quenching models, the ability of a 
galaxy to redden is completely determined by its host halo mass. In the simplest 
secular models, this is completely determined by the galaxy mass. In contrast, 
a merger-driven model depends on both -- mergers will proceed more 
rapidly and efficiently at high $\mgal$ in a given $\mhalo$, and larger $\mhalo$ systems 
represent larger overdensity peaks which are more evolved and more likely to have 
undergone a period of merging (recall, we refer to 
accreting a pair of mass $\sim\mgal$ to fuel a major galaxy merger). 

\begin{figure*}
    \centering
    \figexpand
    \plotone{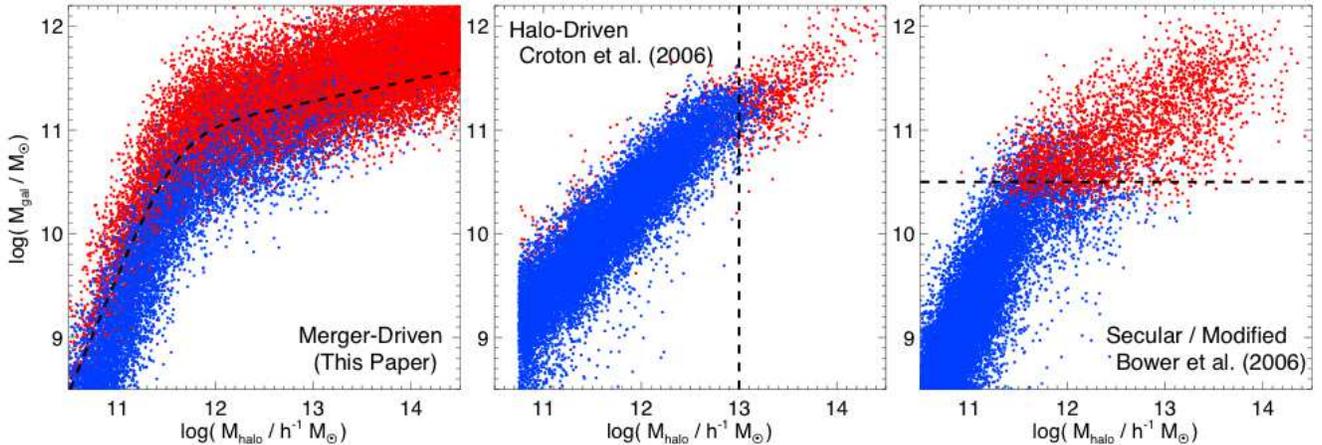}
    \caption{As Figure~\ref{fig:cartoon}, but showing the predictions from 
    full cosmological models (again, for central galaxies only). Galaxies 
    are color-coded by whether or not each model predicts they should 
    be in the blue cloud or red sequence.
    {\em Left:}
    Our full merger model Monte Carlo predictions. 
    {\em Center:} 
    The semi-analytic model of \citet{croton:sam}, which 
    implements a standard halo quenching model (albeit requiring the 
    presence of a relatively massive BH to maintain quenching). Note the 
    apparent relatively low number of massive galaxies/halos owes to 
    the sampling density of the model in its public release. 
    {\em Right:} The modified semi-analytic model of \citet{bower:sam}, 
    as described in \S~\ref{sec:ellipticals:fractions}, 
    where we assume the strong secular (disk instability) mode that 
    dominates the morphological transformation and gas exhaustion of most 
    disks (in the model) also determines whether or not galaxies are quenched. 
    Dashed lines in each qualitatively divide the 
    red and blue populations, as in Figure~\ref{fig:cartoon}. Despite 
    the considerably complexity added to these models, their qualitative 
    behavior in the $\mgal-\mhalo$ plane reflects the key 
    distinctions of each corresponding toy model in Figure~\ref{fig:cartoon}. 
    \label{fig:real.cartoon}}
\end{figure*}

Of course, the cartoon illustration in Figure~\ref{fig:cartoon} ignores some details. 
In many halo quenching models, quenching also requires a massive BH or 
some other feedback mechanism, which implicitly requires a relatively massive 
spheroid and therefore depends to some extent on the 
stellar mass and merger history of the system. In many secular models, galaxy 
structure and disk instability are influenced by halo properties (e.g.\ concentration) 
that vary with halo mass and accretion history. We therefore consider 
a more detailed comparison with state-of-the-art semi-analytic models. 
We extract the results of 
\citet{croton:sam}\footnote{http://www.mpa-garching.mpg.de/NumCos/CR/Download/index.html} 
and \citet{bower:sam}\footnote{http://galaxy-catalogue.dur.ac.uk:8080/galform/}, 
both recent fully cosmological semi-analytic models based on the 
Millenium dark-matter simulation \citep{springel:millenium}. 

The \citet{croton:sam} 
models correspond roughly 
to the halo quenching models described above -- a massive BH is required 
to maintain the hot halo, but development of the hot halo reservoir (upon 
crossing the appropriate halo mass threshold) is 
still effectively the dominant criterion for quenching
\citep[see also e.g.][]{kang:sam,cattaneo:sam,delucia:sam}. 

The \citet{bower:sam} models implement a strong 
disk instability (secular) mode, which dominates black hole growth and 
bulge formation at all redshifts, with 
mergers typically contributing only $\sim0.1\%$ to the spheroid mass budget. 
However, in the model, it is still assumed that cooling can only be halted in a 
quasi-static hot halo, and effectively galaxies are quenched upon crossing the 
appropriate halo mass threshold (like other models, the presence of a moderate-mass BH 
is technically required, but essentially all systems with sufficiently massive halos easily 
host a BH of the necessary mass, even without mergers, owing to the 
disk instability mode of growth). For our purposes, therefore, 
it is effectively equivalent to the \citet{croton:sam} and other 
halo quenching models. But, given the strong secular mode assumed in the model, 
we easily can use it to construct an mock example of a semi-analytic model in which 
secular processes dominate the quenching itself. 

We do so by adopting the 
\citet{bower:sam} model, but instead of using their criterion for 
quenching (namely, the presence of a hot halo), simply assume that systems which 
undergo a sufficiently massive disk instability 
that destroys the entire disk will ``quench.'' 
The disk stability is estimated according to the assumptions of the original model, 
based on e.g.\ disk angular momentum, scale lengths, masses, 
and concentrations. 
We specifically adopt a mass threshold for the instability of $\gtrsim2\times10^{10}\,\msun$ 
(i.e.\ assume systems in which less of the galaxy mass participates in the instability will not 
automatically quench, since almost all galaxies in the model have at least some very 
small mass added to the bulge via instabilities). We choose this value because it 
gives a good match to the total observed mass density of passive galaxies and 
globally-averaged quenched fractions as a function of $\mgal$, but note that our comparisons 
are all qualitatively unchanged regardless of exactly how we choose the 
quenching criterion. We subsequently refer to this as a ``modified \citet{bower:sam}'' model, 
and emphasize that we are not plotting the predictions of the original model 
\citep[which are, for our purposes, equivalent to the predictions of][]{croton:sam}, but using 
it to represent the predictions of a cosmological model for secular evolution, in the case 
where that evolution dominates galaxy quenching. 

We extract 
the $z=0$ predictions of both models, and classify galaxies as either red or blue following 
the criteria of the authors (namely colors $(U-B)>0.8$ being ``red''), although it does 
not change our qualitative comparisons if we adopt a magnitude-dependent 
color limit \citep[although, as noted by][this reveals that high-mass galaxies in both 
models are ``too blue'']{weinmann:group.cat.vs.sam}. We extract these properties 
only for central galaxies -- both semi-analytic models invoke 
alternative physical mechanisms such as ram pressure stripping 
to rapidly redden essentially all satellite galaxies. While there is no doubt this is 
an important mechanism, it has nothing to do with the models we wish to compare, 
and would only confuse the comparison we wish to highlight (and obscure 
the important differences between models). The position of these systems in 
the $\mgal-\mhalo$ space is shown in Figure~\ref{fig:real.cartoon}, in the 
same manner as Figure~\ref{fig:cartoon}. 

To compare to these models in more detail, we construct a 
realistic Monte Carlo population of galaxies of different masses in different mass halos, 
from our merger-driven model. 
Beginning with a small halo at high redshift, 
hosting a (initially) 
disk-dominated galaxy (in the absence of mergers), 
we integrate forward in time. 
The average halo mass accretion history in a 
$\Lambda{\rm CDM}$ universe is well-defined \citep[here we adopt 
the average progenitor mass as a function of time from][]{neistein:natural.downsizing}.
At each point in time, the average mass of a disk galaxy in such a halo can 
be estimated empirically, either from halo occupation models \citep[e.g.,][]{yang:clf,
conroy:monotonic.hod,wang:sdss.hod}, 
adopting the baryonic Tully-Fisher relation \citep[assuming the 
disk circular velocity traces maximal halo circular velocity, e.g.][]{mcgaugh.tf.old,
mcgaugh:tf,belldejong:tf}, or assuming a 
constant baryon fraction in the galaxy. 
We henceforth adopt the baryonic Tully-Fisher expectation for 
$M_{\rm disk}(\mhalo)$, which we assume does not evolve with redshift 
\citep[as suggested by a large number of observations at least to $z\sim1.5$, 
and by some to $z\gtrsim3$;][]{conselice:tf.evolution,
flores:tf.evolution,bell:tf.evolution,kassin:tf.evolution,vandokkum:tf.evolution}, 
but we have tried all three estimators, and find similar results. 
This is not surprising, since, as discussed in \S~\ref{sec:mergers}, 
observations find there is little or no evolution in most general halo occupation 
statistics of star-forming galaxies (i.e.\ average baryonic mass 
hosted by an ``un-quenched'' halo of a given mass) even to $z\sim4$ 
\citep{yan:clf.evolution,heymans:highz.baryon.fractions,
conroy:monotonic.hod,conroy:mhalo-mgal.evol}. 

At each time,
we probabilistically increase the disk mass with the halo mass, such that 
an ensemble of these Monte Carlo simulations always has the appropriate 
mean $M_{\rm disk}(\mhalo)$ and observationally measured scatter about this 
quantity.  Then, we calculate the probability of a major gas-rich merger, 
specifically the probability both that the halo has accreted another halo hosting a 
galaxy of comparable mass (mass ratio $<3:1$) and that the two will merge 
in the given timestep. This calculation is identical to that in \S~\ref{sec:mergers} 
(see \paperone\ for details), 
where the former probability 
has been determined from dark-matter simulations (i.e.\ the probability 
of hosting or accreting a subhalo of the appropriate mass range) and the latter 
is the ratio ${\rm d}t / \tmerger$ (where $\tmerger$ is the merger timescale as 
in \S~\ref{sec:mergers}; we generally adopt the dynamical friction timescale 
in what follows, but our 
results are qualitatively similar regardless of this choice). 
Based on this probability, it is randomly determined whether or 
not the galaxies merge. If so, the final stellar mass is just the sum of the two 
pre-merger baryonic masses, and we assume zero further growth through 
star formation (although growth via dry mergers is allowed). 

We technically
integrate this model only from $z\sim10$ to $z=0$ (or where $\mhalo>10^{9}\,\msun$), 
but find the results are reasonably converged with respect to this choice (although 
in principle every halo may have a major merger if we integrated to 
infinite redshift or $\mhalo=0$, these mergers are meaningless for our purposes 
as there is no significant galaxy formed inside the halo). 
Running a large sample of Monte Carlo realizations for each $\mhalo$, we 
obtain a bivariate $z=0$ distribution of early and late-type galaxies in 
$\mgal$ and $\mhalo$ which reflects our models. The resulting predictions 
are shown in Figure~\ref{fig:real.cartoon}. 

Although this in some sense serves as a crude toy semi-analytic model, 
we adopt this approach specifically to minimize the uncertainty owing to choices 
such as the modeling of star formation and accretion in galactic disks. Instead, we 
adopt as much as possible in a purely empirical fashion, to isolate 
the predictions of a merger-driven quenching model (and not confuse these 
with degeneracies in modeling disk formation). Since mergers will efficiently convert 
gas to stars, and their gravitational processes are not changed by the ratio of 
gas to stellar mass, our results are also entirely independent of the star formation 
histories in the disks -- we only need to inform our predictions with a rough 
estimate of the masses of disks hosted by halos of a given $\mhalo$. Ultimately, 
adding the complications of our Monte Carlo tests allows us to construct a comparison to the 
\citet{croton:sam} and modified \citet{bower:sam} models, but yields a qualitatively similar 
result to our naive cartoon expectation in Figure~\ref{fig:cartoon}. 
(Note that there are some differences in the low-mass star-forming galaxies 
between the various models, owing to their treatment of star formation, but 
this is unimportant for any of our conclusions.) 

\begin{figure}
    \centering
    \figexpand
    \plotter{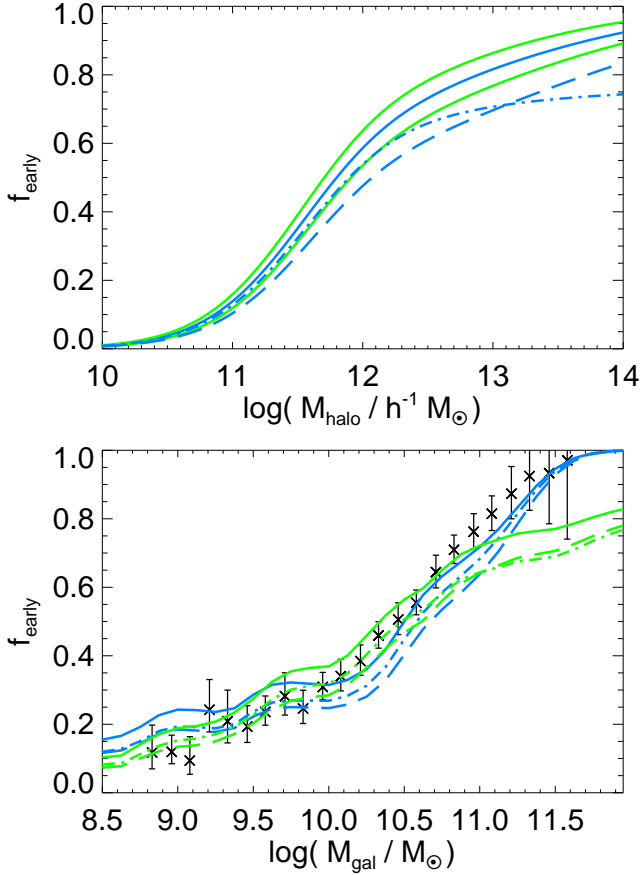}
    \caption{{\em Top:} Local fraction of red/early-type (major merger remnant) 
    central galaxies 
    as a function of halo mass, from our prediction in Figure~\ref{fig:real.cartoon}. 
    Linestyles adopt different estimations of the merger rate, as in Figure~\ref{fig:local.mf}.
    Solid blue line shows the mean fraction, upper and lower green lines the fraction in 
    the higher and lower stellar mass halves at each $\mhalo$, respectively. 
    {\em Bottom:} Same, as a function of galaxy stellar mass. Green lines in this 
    case are as the blue lines, but adopt a different halo occupation fit
    (as the dashed black lines in Figure~\ref{fig:local.mf}).
    Black squares show the observed early-type galaxy 
    fraction as a function of mass from \citet{bell:mfs}. 
    \label{fig:red.frac.summary}}
\end{figure}
\begin{figure}
    \centering
    \figexpand
    \plotone{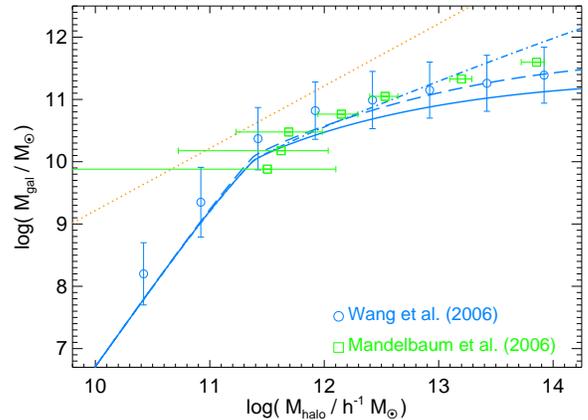}
    \caption{Mean central galaxy baryonic 
    mass as a function of halo mass (blue lines, as in Figure~\ref{fig:local.mf}), from our prediction 
    in Figure~\ref{fig:real.cartoon}. Dotted orange line corresponds to the universal 
    baryon fraction. 
    Points show the observationally estimated mean 
    central galaxy stellar mass as a function of halo mass 
    from \citet{wang:sdss.hod} (HOD fitting) and \citet{mandelbaum:mhalo} (weak lensing).
    The merger-driven quenching model 
    naturally predicts the red fraction as a function of mass and the turnover in 
    the $\mgal(\mhalo)$ relation (equivalently, turnover in galaxy $M/L$ ratios above 
    $\sim\mstar$), without any input parameters describing a preferred mass scale. 
    \label{fig:mgal.mean.vs.mhalo}}
\end{figure}

Quantitatively, we can now integrate the results of Figure~\ref{fig:real.cartoon} 
and predict the red (i.e.\ merger remnant) fraction as a bivariate 
function of $\mgal$ and $\mhalo$; Figure~\ref{fig:red.frac.summary} shows this. 
In order to represent the real observations, we add the appropriate 
observational errors in both $\mgal$ and $\mhalo$ ($\sigma_{\mgal}\approx0.2$\,dex, 
$\sigma_{\mhalo}\approx0.4$\,dex), for both our model and the \citet{croton:sam,bower:sam} 
models. This does not qualitatively change any of the results, but does smooth some of 
the dependencies (and tends to remove unphysical features in the models 
caused by undersampling). 

In a global sense, the trends appear to be reasonably accurate -- they agree well 
with the observed fraction of red galaxies as a function of galaxy stellar mass 
\citep{bell:mfs}. Figure~\ref{fig:mgal.mean.vs.mhalo} compares the 
mean $\mgal$ at each $\mhalo$ predicted from this model. Quenching 
associated with  
major mergers naturally predicts the turnover in $\mgal(\mhalo)$ around 
$\mgal\sim10^{11}\,\msun$. We emphasize that there are no parameters in 
our model which have been tuned or otherwise adjusted to give this 
result -- unlike halo quenching models which empirically adopt 
a specific quenching mass, we have no input parameter which 
fixes this mass. Rather, the turnover arises self-consistently, as the result of 
major, gas-rich mergers first becoming efficient at these masses, and subsequent 
star formation being quenched. 

\begin{figure*}
    \centering
    \plotone{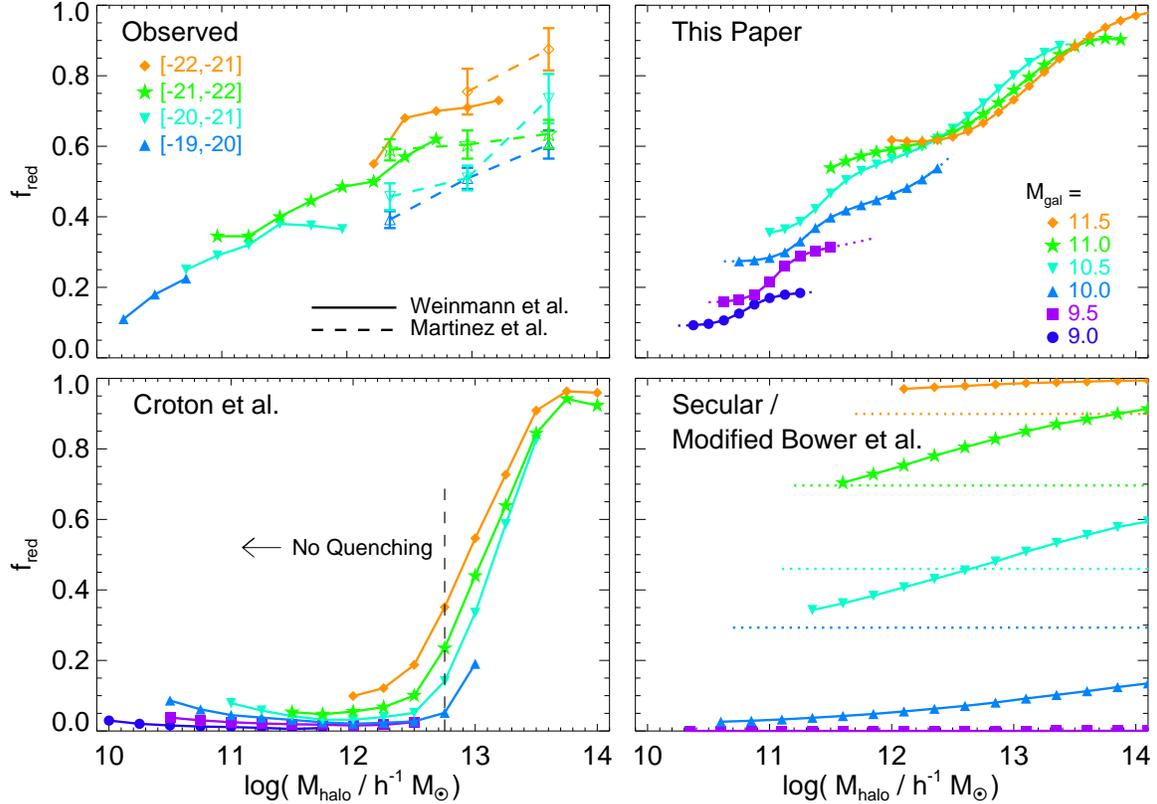}
    \caption{Red/early-type fraction $f_{\rm red}$ (of {\em central galaxies only})
    as a bivariate function of stellar mass/luminosity and 
    halo mass/local environment. 
    We specifically exclude satellites, as they tend to be uniformly red (making the predicted 
    red fractions degenerate between central galaxy quenching mechanisms and 
    the satellite fraction as a function of $\mhalo$ and $\mgal$). 
    {\em Top Left:} Observed $f_{\rm red}$ of central galaxies as a function of host 
    halo mass (estimated from matching group catalogues to halo mass functions) 
    in bins of galaxy $r$-band magnitude, from \citet[][solid line, filled points]{weinmann:obs.hod} and 
    \citet[][dashed line, open points]{martinez:redfrac.groups,martinez:redfrac.evol}.
    (Note there appears to be some small fraction of massive galaxies in small halos in each panel: this 
    owes to scatter in the halo and stellar mass estimators, but has no effect on the conclusions.)
    {\em Top Right:} Predicted $f_{\rm red}$ of central galaxies 
    from our merger model, as a function of halo mass in bins of galaxy stellar mass, as labeled
    (bins of a given color/style roughly correspond to the observed $r$-band absolute 
    magnitude ranges of the same color/style). 
    {\em Bottom Left:} Same, from the \citet{croton:sam} halo quenching model. We 
    qualitatively label the quenching halo mass, which separates uniformly low 
    and uniformly high $f_{\rm red}$ in this model.
    {\em Bottom Right:} Same, from the modified \citet{bower:sam} secular model. Dotted lines 
    show the $\mhalo$-independent red fraction for 
    each $\mgal$ if the model were strictly dependent on only $\mgal$ (the model 
    treats satellites and central galaxies differently, so the normalizations of these 
    $f_{\rm red}$ estimates does not agree at low $\mgal$). 
    The behavior of the three models is qualitatively different, as in Figure~\ref{fig:real.cartoon}, 
    with a merger model predicting a joint dependence on $\mgal$ and $\mhalo$ 
    distinct from the halo quenching or secular models. 
    \label{fig:red.frac.bivar}}
\end{figure*}

We now examine the predicted red fractions in greater detail, by breaking them 
down as a bivariate function of both $\mgal$ and $\mhalo$. Figure~\ref{fig:red.frac.bivar} 
shows this, for each of the three models as in Figure~\ref{fig:real.cartoon}, 
and several observational determinations. Specifically, we calculate the 
red/early-type fractions predicted as a function of $\mhalo$, in bins of 
galaxy stellar mass $\mgal$. 
Note that a detailed quantitative comparison with the observations 
is difficult and beyond the scope of this paper, as 
the exact absolute values of $f_{\rm red}$ depend sensitively on the selection 
method and conversion between group properties and halo mass
\citep[see, e.g.][]{cooper:z1.color.density}. But the qualitative 
trends are robust to these effects \citep[see e.g.][]{weinmann:group.cat.vs.sam,
cooper:color.density.evol,cooper:z1.color.density}.
For all the model predictions and the observational 
analogues, we consider only the red fraction of central galaxies. In most models, 
satellite galaxies are uniformly (or close to uniformly) red, so considering the 
total (central+satellite) red fraction mixes the consequences of the 
physics causing quenching (what makes central galaxies red) with the 
estimated satellite fraction as a function of $\mgal$ and $\mhalo$ (which, while 
importantly informing models, contains no information about the physics of 
central galaxy quenching). 

From the observed group catalogues of 
\citet{weinmann:obs.hod} and 
\citet{martinez:redfrac.groups,martinez:redfrac.evol}, which consider the 
same (again, for central galaxies only), there are a few important qualitative trends. 
These include: (1) a strong dependence of red fraction on halo mass, but
(2) a significant residual dependence on galaxy mass/luminosity, (3) 
a lack of any sharp characteristic scale in $\mhalo$, (4) a relatively high 
red fraction ($f_{\rm red}\gtrsim0.5$) for the most massive/luminous systems even 
at low halo masses ($\mhalo\lesssim10^{12}\,\msun$), and 
(5) a similar, relatively high red fraction ($f_{\rm red}\gtrsim0.5$) 
for the least massive/luminous systems at high halo masses 
($\mhalo\gtrsim10^{13}\,\msun$).

In contrast to the observed trends, 
the \citet{croton:sam} model is, as expected, similar to a pure halo quenching 
model -- there is a sharp transition from uniformly low red fractions ($f_{\rm red}\lesssim0.1$) 
below the halo quenching mass ($\sim$a few $10^{12}\,\msun$) to 
uniformly high red fractions above this halo mass, with a weak residual dependence 
on galaxy mass. The low red fraction at small halo masses also forces these 
models to assume a high red satellite fraction at these masses 
(in order to match the global red galaxy mass functions), in disagreement 
with observations \citep{weinmann:obs.hod}. 

The public (original) version of the \citet{bower:sam} model yields an essentially 
identical prediction to the \citet{croton:sam} model in this space, as 
the development of a hot halo is assumed to be the key criterion for quenching. 
The modified \citet{bower:sam} model which we consider, on the other hand, 
is quite similar to a pure secular model (as expected), 
with $f_{\rm red}$ nearly independent of $\mhalo$ at each $\mgal$. 
There is some weak dependence, because galaxies living in high-mass halos 
tend to have earlier formation times, meaning that their progenitor disks 
were more compact and therefore (according to the model assumptions) 
more prone to massive instabilities, but 
the primary dependence of $f_{\rm red}$ is clearly on galaxy mass.

Neither of these 
predictions agrees qualitatively with the observations. 
The prediction from 
our merger model, however, matches these features -- the dependence 
on $\mhalo$ is stronger than that in the modified \citet{bower:sam} model 
(or a ``toy'' secular model) in considerably 
better agreement with the observations, but the residual dependence on $\mgal$ 
is stronger than that in \citet{croton:sam}. There is no sharp transition at 
some specific $\mhalo$, and the red fraction of massive systems 
remains relatively high at lower $\mhalo$, in contrast to the \citet{croton:sam} 
predictions. However, there is still a significant dependence on halo mass, 
and low stellar mass systems do become red at large $\mhalo$, 
in contrast to the secular/modified \citet{bower:sam} model predictions.

\begin{figure}
    \centering
    \figexpand
    \plotone{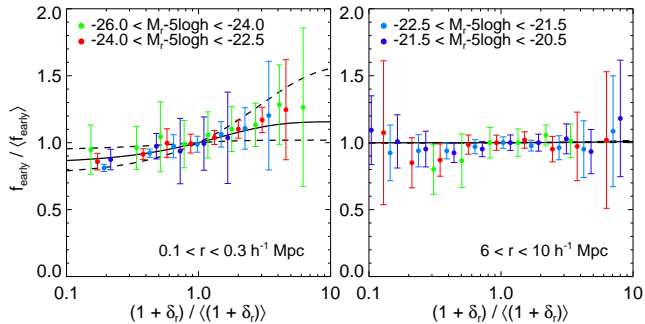}
    \caption{Dependence of red fraction on density at small scales 
    ({\em left}) and large scales ({\em right}), at fixed 
    halo mass (i.e.\ considering $f_{\rm red}/\langle f_{\rm red} \rangle$ 
    versus density $(1+\delta_{r})/\langle (1+\delta_{r}) \rangle$ at fixed $\mhalo$). 
    Points show the observations from \citet{blanton:smallscale.env}, 
    for SDSS groups with different total group luminosities (as labeled; this should 
    be a good proxy for total group halo mass). Lines show our prediction, 
    which has an increasing $f_{\rm early}$ with overdensity on small scales, 
    as a local galaxy overdensity implies an increased probability of major mergers. 
    Solid line is for $z=0$, $\mhalo=\mstar\approx1.5\times10^{12}\,h^{-1}\,\msun$ halos, 
    dashed lines show how this changes for more massive ($\mhalo\sim10^{15}\,h^{-1}\,\msun$, 
    shallower dependence on density) and less massive 
    ($\mhalo\sim10^{10}\,h^{-1}\,\msun$, 
    steeper dependence on density) halos. The merger-driven quenching hypothesis naturally 
    explains a dependence on small-scale overdensity, similar to that observed. 
    \label{fig:fred.density}}
\end{figure}

Observationally, when red fractions are quantified as a function of quantities 
such as galaxy density $\rho$ or surface density $\Sigma$, there is some 
ambiguity in what these quantities represent. To lowest order, they serve 
as tracers of halo mass and are 
directly comparable to predictions such as those in 
Figure~\ref{fig:red.frac.bivar}. However, \citet{baldry06:redfrac.vs.m.env} and 
others have suggested that the trends 
in $f_{\rm red}$ with these quantities argue for some level of dependence 
on environment, even after accounting for the 
primary dependence on halo mass. 

In greater detail, 
\citet{blanton:smallscale.env} investigate this possibility by determining 
$f_{\rm red}$ in SDSS groups as a function of local density ($1+\delta_{r}$; 
defined as the galaxy number density within a radius $r$ relative 
to the mean number density in that radius) on 
various scales, in narrow bins of total group luminosity \citep[which 
should be a good proxy for group halo mass, see][]
{yang:group.finder,vandenbosch:concordance.hod}. 
At fixed total group luminosity (roughly equivalent to fixed group parent halo 
mass), they find no 
evidence for an additional dependence of red fraction on large-scale 
environment, measured at projected radii $6 < r < 10$ 
and $0.3 < r < 1\,h^{-1}\,{\rm Mpc}$. 
However, at small radii 
$0.1 < r < 0.3\,h^{-1}\,{\rm Mpc}$ they find a significant dependence of 
$f_{\rm red}$ on density for all group luminosities (halo masses) 
which they consider. A similar result is found by 
\citet{park:redfrac} and \citet{kauffmann:sf.vs.env}. 

In a halo quenching or secular model, this 
is difficult to explain, as quenching depends only on either halo mass or 
internal galaxy properties, respectively. However, as we have discussed 
for both ongoing/recent galaxy mergers and 
quasars in \paperone, mergers are more likely to occur in regions 
with galaxy overdensities on very small scales. The bias to increasing 
fractions of merger remnants with increasing small scale density in 
\paperone\ (see Figures~7 \&\ 17 in that paper) directly translates to a prediction 
for the dependence of $f_{\rm red}$ on small scale overdensity, which we 
show in Figure~\ref{fig:fred.density}. The merger hypothesis provides a natural 
explanation for the observed dependence on small-scale overdensities. 

We caution, however, that this explanation is not unique. If satellites 
are preferentially red, then a simple autocorrelation function or dependence of 
overall red fraction on density \citep[such as that observed by][]{blanton:smallscale.env} of 
red galaxies will see a similar effect (with the excess on small scales reflecting 
the abundance of satellite galaxies). Furthermore, mergers (by definition) consume 
some of the galaxies that (initially) define the small-scale overdensity, so it is not clear 
how much of this effect might be wiped out by the mergers themselves. In other words, 
seeing this effect weakly does not necessarily argue against a merger-driven model 
for quenching, and theoretical study in cosmological simulations is needed for 
more detailed predictions.

A more rigorous test of this would 
be to compare the cross-correlation of central red galaxies with 
all other galaxies, i.e.\ to more directly test whether central red galaxies preferentially 
live in regions of small-scale galaxy overdensity. Early analysis along these 
lines does suggest a similar conclusion \citep{masjedi:cross.correlations}, 
in agreement with these 
predictions (and difficult to reconcile in models where quenching is a pure 
function of galaxy or halo mass). 

\subsection{Redshift Evolution of Quenched Fractions and Color-Morphology-Density Relations}
\label{sec:ellipticals:evolution}

\begin{figure}
    \centering
    \figexpand
    \plotone{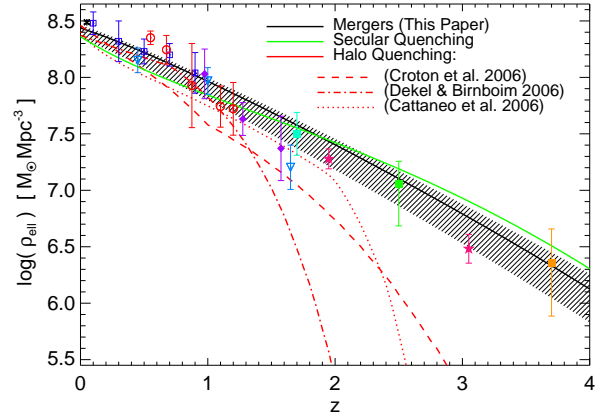}
    \caption{Integrated stellar mass density of red or early-type galaxies as a function of redshift. 
    The predictions from a merger quenching model (with shaded range reflecting the range of 
    predictions from our different adopted models) 
    and observations are as in Figure~\ref{fig:red.mass.density}. We compare the predictions of 
    a pure secular model (based on the observed total mass function in \citet{fontana:highz.mfs}, 
    where we assume the red fraction as a function of galaxy mass is identical to that 
    at $z=0$), 
    and various halo quenching models (where we assume all systems above the quoted 
    halo quenching masses in our halo occupation model are quenched). 
    In order to form massive galaxies at high redshift, 
    the traditional halo quenching models must allow cold accretion flows in all but the most massive 
    halos at $z\gtrsim2-3$, yielding almost no red galaxies at these redshifts. In contrast, by allowing 
    reddening to occur in a range of halo masses (which might otherwise continue accreting), 
    the merger and secular model produce a sufficient density of passive galaxies at high 
    redshift. 
    \label{fig:red.mass.density.models}}
\end{figure}
We next examine the redshift evolution of the trends in red galaxy 
fraction with stellar and/or halo mass. 
First, we return to our prediction for the mass density of passive systems as a function 
of redshift (Figure~\ref{fig:red.mass.density}). As noted there, a merger quenching 
model predicts a (relatively high) 
density of passively evolving, quenched systems in good agreement 
with that observed at high redshifts $z\gtrsim2-3$. 
This is in strong contrast to many pure hot-halo quenching models, in 
which cold accretion within a hot halo persists at high redshifts $z\gtrsim2$. 
Figure~\ref{fig:red.mass.density.models} contrasts the merger-driven prediction 
with that from several halo quenching models \citep{dekelbirnboim:mquench,
cattaneo:sam,croton:sam}, where we assume 
(since we are ultimately just making a qualitative comparison) that 
all systems above their quoted quenching halo mass thresholds are red/passive 
(but our explicit comparison with these models in \S~\ref{sec:ellipticals:fractions} 
suggests this is actually a good approximation to their predictions). Note that 
for the \citet{cattaneo:sam}, this is a comparison with the halo quenching-only 
version of their model (since they also consider a model in which, like ours herein, 
major merger remnants are automatically quenched). We then use our 
halo occupation model to determine the ``red'' mass density. 

The result is 
similar in each halo quenching model: above $z\sim2-3$, the density of passive galaxies plummets. 
In detail, for example, 
the \citet{dekelbirnboim:mquench} derivation of the hot halo quenching mass 
predicts that at $z\sim3.5$, cold flows continue within all $\mhalo \lesssim10^{14}\,\msun$ 
halos, which allows only a completely negligible maximum 
red galaxy mass density (since such halos are extremely rare at high redshifts). 
As demonstrated by \citet{cattaneo:sam}, introducing these ``cold flows in hot halos'' is 
necessary for these models to match the overall density of massive galaxies 
and cosmic star formation rate density at 
high redshift. This owes to the steep step-function transition from unquenched to 
quenched systems around the quenching mass 
in such models (see Figure~\ref{fig:red.frac.bivar}) -- 
if the quenching mass is lowered (to make for more red galaxies), then the models 
will quench systems too early, and not form any 
high-redshift massive galaxies in the first place. 

In contrast, a merger driven model 
is able to predict that the appropriate fraction of these massive, high 
redshift galaxies are passive. This is because it allows for reddening to 
be somewhat uncoupled from halo mass; i.e.\ systems in massive halos might 
generally continue to accrete, but some fraction can redden and build up 
sufficient mass density of passive galaxies (without reddening all systems of these 
masses and destroying the ability to make massive galaxies in short cosmic times). 

In addition, we compare the simplest secular model (where we just adopt our halo occupation
model and assume the red fraction as a function of galaxy mass is identical to that 
at $z=0$; as in \S~\ref{sec:ellipticals:fractions}), and (for the same reasons) 
find that it is also able to reproduce the observations. This should not be surprising, 
since in this toy model, matching the overall galaxy mass density is 
implicit, and the red fractions are as high as they are at $z=0$. However, we 
will show that there are other aspects in which the redshift evolution of 
red fractions predicted by a secular model are in conflict with the observations. 

Interestingly, simulations suggest that hot halos often develop at lower 
masses than the halo quenching models require 
\citep[$\sim10^{11.5}\,\msun$; see][]{keres:hot.halos} -- in other words, the possibility that 
such halos are {\em necessary} for quenching is viable, since this mass threshold 
allows for the possibility of sufficient populations of passive, high-redshift galaxies. 
However, the idea that such halos are {\em sufficient} for quenching (as 
is effectively true in the halo quenching models) is not viable, 
since it prevents the formation/growth of galaxies beyond this mass threshold 
and cannot form sufficient numbers of massive galaxies nor yield a sufficiently 
high global star formation rate at high redshift. 

\begin{figure}
    \centering
    \plotter{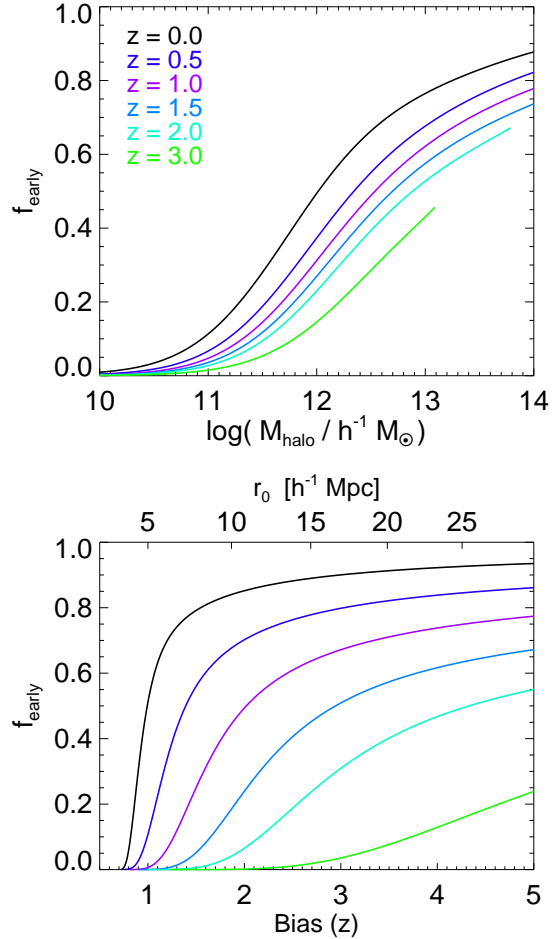}
    \caption{{\em Top:} Evolution in the predicted early-type fraction as a function of 
    halo mass (see Figure~\ref{fig:red.frac.summary}) with redshift (here for 
    just our standard model).
    {\em Bottom:} Same, but as a function of the halo clustering 
    amplitude (calculated for each $\mhalo$, $z$) or (rough) equivalent 
    correlation length $r_{0}$. The figure should be interpreted as reflecting 
    the scales on which a color-density or morphology-density relation 
    will manifest at different redshifts. At high redshifts, the highest-density 
    regions (e.g.\ $b\gtrsim2$, $r_{0}\gtrsim10$ at $z\sim2$) 
    will have built up a color-density relation, while at low redshifts 
    the color-density relation will have built up even in field populations ($b\sim1$). 
    \label{fig:fred.z.pred}}
\end{figure}
Figure~\ref{fig:fred.z.pred} shows the mean predicted early-type fraction 
in our merger-driven model as a 
function of halo mass (as in Figure~\ref{fig:red.frac.summary}) at 
different redshifts. We also plot this as a function of 
the estimated clustering amplitude for each $\mhalo$ (at each redshift), and 
the (approximate) corresponding comoving correlation length $r_{0}$ 
(where we define $r_{0} \equiv 5\,h^{-1}\,{\rm Mpc}\,[b(z)]^{2/\gamma}$, 
where $\gamma\approx1.8$ as measured locally). At high redshifts,
systems in only the most massive halos (i.e.\ most extreme overdensities) 
have sufficiently rapid merger rates that they will have built up large  
red fractions. In terms of $\mhalo$, the evolution appears relatively 
weak, but in terms of the clustering amplitude, it is more 
obvious (this owes to massive halos being rarer
and corresponding to significantly higher-density 
peaks at high redshifts). 

\begin{figure}
    \centering
    \figexpand
    \plotone{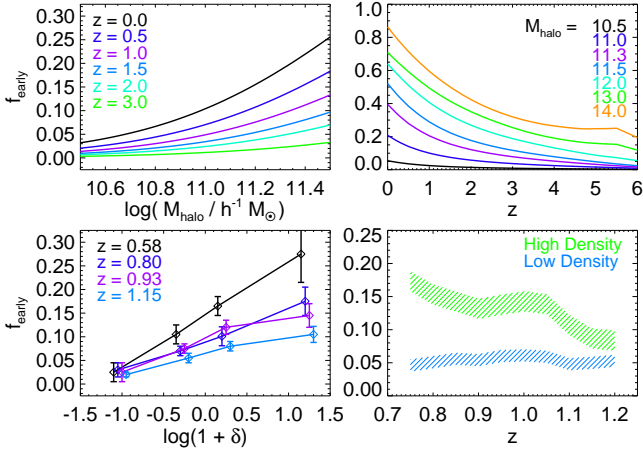}
    \caption{As Figure~\ref{fig:fred.z.pred}, but comparing with 
    observations. {\em Top Left:} Predicted red fraction versus 
    halo mass and redshift, for typical halo masses (near $\mstar$) at $z\sim0-1$. 
    {\em Bottom Left:} Observed red fraction verses density (which should 
    trace $\mhalo$, to lowest order) at different redshifts from \citet{cooper:color.density.evol}. 
    The predicted and observed relations flatten with redshift in a similar manner. 
    {\em Top Right:} Predicted red fraction versus redshift, for different halo masses 
    ($\log(\mhalo/h^{-1}\,\msun) = 10.0$, etc., as labeled). The most massive halos 
    are in high-density regions which evolve most rapidly at high redshifts. 
    {\em Bottom Right:} The observed red fraction versus redshift 
    from \citet{cooper:color.density.evol} 
    for the highest-density and 
    lowest-density $1/3$ of systems at each redshift (shaded ranges). 
    An exact quantitative comparison is sensitive to color selections and 
    definitions of environment, but the qualitative trends are similar. 
    \label{fig:fred.z.obs}}
\end{figure}

Figure~\ref{fig:fred.z.obs} compares these predictions to 
recent observations from \citet{cooper:color.density.evol}. The same trends are 
evident for both our predictions and the observations. Looking at typical 
halos, expected in average regions of the universe, the color-density 
relation (more specifically, the dependence of $f_{\rm early}$ on halo 
mass) nearly vanishes by $z\sim1.5$. Similar trends have been 
observed by \citet{gerke:blue.frac.evol} and \citet{nuijten:color.density.evol}. 
This does not mean that there is no 
such trend -- it does not become prominent until higher-density, 
rare high-halo mass peaks are probed, as in Figure~\ref{fig:fred.z.pred}. 
This is also observed -- for the most massive, red galaxies at $z\sim2-4$, 
a color-density relation is seen for dense environments with 
$r_{0}\gtrsim10\,h^{-1}\,{\rm Mpc}$, similar to our predictions. 
Again, we emphasize that a detailed quantitative comparison is outside the 
scope of this paper, as it is sensitive to the galaxy and color selection method 
and the exact definition of galaxy environments \citep{cooper:color.density.evol}, 
but the qualitative trends should be robust. We also caution that the observations 
(presently) mix central and satellite systems, although above moderate 
$\sim\lstar$ luminosities, satellite systems should only be a small fraction of the 
observed populations. 

\begin{figure}
    \centering
    \figexpand
    \plotone{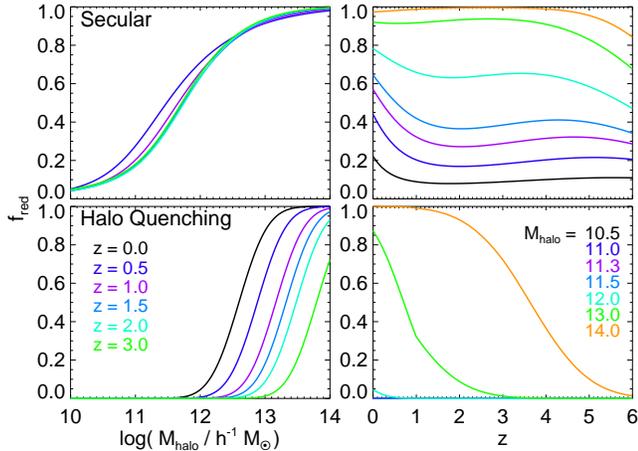}
    \caption{Red or quenched fraction as a function of halo mass and redshift 
    (as Figures~\ref{fig:fred.z.pred} \&\ \ref{fig:fred.z.obs}), for a pure 
    secular model ({\em top}) and a pure halo quenching model ({\em bottom}). 
    Because the average total mass of a disk in a halo of a given mass 
    evolves weakly with redshift, there is little evolution in the 
    color-density (morphology-density) relations in the secular model, 
    in contrast to the observations.
    \label{fig:fred.z.models}}
\end{figure}

In Figure~\ref{fig:fred.z.models} we
compare the evolution in the red fraction predicted as a function 
of halo mass and redshift from our merger-driven model to that 
predicted in a secular or halo quenching model. For the secular 
model, we assume that the red fraction is a pure function of galaxy 
mass (calibrated at $z=0$ as in Figure~\ref{fig:red.frac.bivar}) -- 
so the difference in red fraction as a function of halo mass reflects 
only evolution in the typical stellar masses hosted in halos of a given 
mass. As we have noted (see \S~\ref{sec:mergers}), this evolves 
relatively weakly, and as a consequence there is little 
evolution in the red fraction as a function of halo mass in secular models. 
It is clear in Figure~\ref{fig:fred.z.models} that (in the secular model) 
the color-density relation for 
``typical'' $\sim10^{11}-10^{12}\,h^{-1}\,\msun$ mass halos does 
not vanish at $z\sim1.5$ as is observed. Although there might be 
some apparent evolution (since, as noted in Figure~\ref{fig:fred.z.pred}, 
the clustering of halos of fixed mass will vary with redshift), 
there is little evolution in the color-density relations in 
terms of halo mass in this model, and it is not obvious how 
any fundamentally secular-dominated model can avoid this
prediction. 

Interestingly, 
\citet{cassata:blue.spheroid.frac} 
observationally estimate that the pseudobulge fraction as a function of 
redshift may exhibit this behavior (being essentially constant for 
a given galaxy/halo mass), as expected if these are formed via secular 
mechanisms, and \citet{sheth:bar.frac.evol} find 
similar results for the evolution in disk bar fractions. 
But \citet{cassata:blue.spheroid.frac} (and others) estimate that this pseudobulge population 
accounts for only $\lesssim5\%$ of massive spheroids. 
Future observations and direct 
comparison with e.g.\ the model of \citet{bower:sam} can 
place stronger constraints on these distinctions, but 
even at present, the observations suggesting 
evolution in the color-density relations at $z\gtrsim1$ 
\citep{cooper:color.density.evol,gerke:blue.frac.evol,nuijten:color.density.evol} 
appear to contradict the basic prediction of a model in which secular processes 
dominate red galaxy formation.

We also consider the predictions from a halo quenching model 
in Figure~\ref{fig:fred.z.models}. Specifically, we adopt the 
red fraction as a function of halo mass from \citet{croton:sam} at 
$z=0$ (a near step-function rise near the halo quenching mass), and 
renormalize it by the evolution in the halo quenching mass 
with redshift (i.e.\ shift the step function to whatever halo mass corresponds 
to the model halo quenching mass at that redshift). The predictions are 
somewhat different from those of our merger-driven model, but in this 
case that owes mostly to the difference in red fraction as a function of 
halo mass at fixed redshift (as in Figure~\ref{fig:red.frac.bivar}). 
Future observations of the red fraction as a bivariate function of 
galaxy and halo mass at different redshifts can break these degeneracies, 
but for now we note that the evolution is also qualitatively 
consistent with the observations -- in both this model and the merger-driven 
model, the field color-density relation begins to disappear around $z\gtrsim1$, 
with the buildup of the color-density relation occurring at higher masses (higher density 
environments) at higher redshifts. 

\subsection{The Role of Dissipationless or ``Dry'' Mergers}
\label{sec:ellipticals:dry}

\begin{figure}
    \centering
    \figexpand
    \plotone{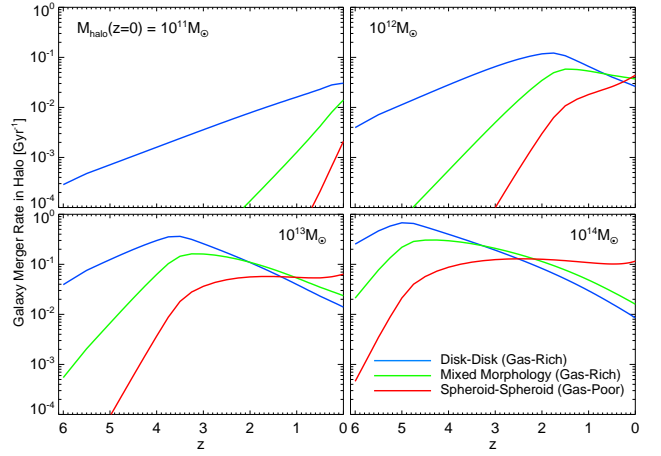}
    \caption{Average predicted major merger rate for central galaxies in 
    halos of a given $z=0$ mass (as labeled). Lines show the average 
    rate of disk-disk (gas-rich), mixed morphology (disk-spheroid, but also 
    gas-rich), and 
    spheroid-spheroid (gas-poor, i.e.\ dissipationless or ``dry'') mergers. 
    In an integrated sense, mixed-morphology mergers are a
    relatively small contribution to global merger rates.
    Dissipationless mergers, however, 
    become dominant in the massive ($\mhalo>10^{11}\,\msun$) systems at 
    late times (once most systems of the given mass have already undergone 
    at least one major merger). 
    \label{fig:merger.types.z}}
\end{figure}

Our model directly yields the major merger history of a given 
galaxy population. We therefore briefly quantify this as a function of 
galaxy properties, decomposing the types of 
mergers in comparison to various observations. 
Figure~\ref{fig:merger.types.z}, for example, shows the average 
major merger rate predicted by our model (specifically our Monte Carlo 
realization described in \S~\ref{sec:ellipticals:fractions}) for central galaxies of a 
given mass, decomposed into the average rates of 
disk-disk mergers (i.e.\ two systems which have not yet 
undergone a major merger), mixed-morphology mergers, 
and spheroid-spheroid mergers (i.e.\ two systems 
which have both undergone previous major mergers). 

Once 
the fraction of systems which 
have undergone at least one major merger 
at a given $\mgal$ and $\mhalo$ becomes large, spheroid-spheroid 
mergers will naturally become the dominant type of merger. In 
other words, merger efficiencies are not especially sensitive to 
galaxy types (at fixed mass), and so reflect the abundance of 
merged or un-merged systems.
We note that only this category, the spheroid-spheroid mergers, 
will be (in our model) gas-poor, dissipationless or ``dry'' mergers.
We show the results just from our default model, but note that they 
are qualitatively similar regardless of our choices of 
halo occupation models, subhalo mass functions, or 
merger timescales. 

The rate of dry mergers in massive systems is consistent with 
observational estimates \citep{bell:dry.mergers,vandokkum:dry.mergers} 
of roughly $\sim0.5-1$ dry mergers per massive elliptical since $z\sim1$ 
(i.e.\ $\sim0.1\,$Gyr$^{-1}$ in $\mhalo\gtrsim10^{13}\,\msun$ halos at $z<1$). 
Although briefly important in the transition between the dominance of 
gas-rich and gas-poor mergers, mixed morphology mergers are 
an intermediate phenomenon -- 
most galaxies that have undergone only their 
initial, gas-rich, spheroid forming major merger were produced in 
disk-disk mergers, and the evolution of massive systems with 
multiple mergers is dominated (at late times) by spheroid-spheroid 
mergers. Note that for most of our predictions above, only the 
fact that the merger remnant is quenched is important, although 
the morphologies of the progenitors can change the ``type'' of merger.

\begin{figure}
    \centering
    \figexpand
    \plotone{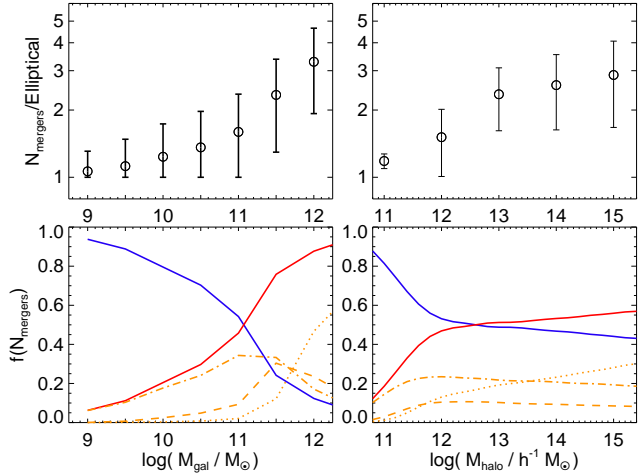}
    \caption{{\em Top:} Average number (with $\sim1\,\sigma$ range 
    shown as error bars) of major mergers 
    in the history of a $z=0$ early-type galaxy (i.e.\ galaxy with 
    at least one major merger in the past) as a function of 
    galaxy stellar mass ({\em left}) or host halo mass ({\em right}). 
    {\em Bottom:} Fraction of early-type galaxies 
    (as a function of galaxy stellar mass [{\em left}] or 
    host halo mass [{\em right}]) which have 
    undergone just their one, initial (gas-rich) spheroid-forming 
    major merger (blue), or more than one major merger (red). 
    Orange lines decompose the red line into systems which have 
    undergone exactly 2 major mergers (i.e.\ 1 gas-rich, and 
    generally 1 spheroid-spheroid or ``dry'' merger; dot-dashed), 3 
    major mergers (2 dry mergers; dashed), and 
    $\ge4$ major mergers ($\ge3$ dry mergers; dotted). 
    \label{fig:dry.int}}
\end{figure}

Integrating these rates to $z=0$, Figure~\ref{fig:dry.int} shows the 
mean number of major mergers (and fraction of 
spheroids with a given number of previous major mergers) 
as a function of spheroid or host halo mass.  We show this 
only for spheroids, since (by definition) the fraction of late-type 
systems (i.e.\ systems which have not undergone a major 
merger) is identical to our blue fractions in \S~\ref{sec:ellipticals:fractions}. 
There is a general trend for more massive systems to 
have experienced a larger number of major mergers, as expected since 
such systems form earlier and in more dense environments. 
The trend is somewhat steeper as a function of galaxy mass than 
as a function of halo mass -- this is also expected, since at a given 
$\mhalo$, a larger number of mergers builds a more massive galaxy, 
steepening the trend in number of mergers as a function of $\mgal$. 

The number of mergers as a function of mass is similar to 
that predicted by various semi-analytic models 
\citep[e.g.,][]{khochfar:sam,croton:sam,kang:sam} and relatable to (although not 
identical to, owing to the difference in definitions) the 
``effective number of progenitors'' in \citet{delucia:ell.formation}. This is expected, 
as mergers are a dynamical inevitability. There might be some 
differences owing to various prescriptions for merger 
timescales or different populations of galaxies in a given halo, 
but the general results of Figure~\ref{fig:dry.int} are robust.
The nature of these mergers, however (whether they are, for example, 
gas-rich or gas-poor), does depend on the model.

It is well-established that there is a general dichotomy in the properties of elliptical 
galaxies \citep[e.g.][]{bender:ell.kinematics,bender:ell.kinematics.a4,
kormendy77:correlations,kormendy:wetvsdry,lauer:bimodal.profiles}. 
Whether or not the division is strict \citep[see e.g.][]{ferrarese:profiles}, 
the most massive ellipticals tend to be slowly rotating, anisotropic systems with 
boxy isophotal shapes and central core profile deviations from a pure Sersic 
profile. Less massive ellipticals, including most of the $\sim\lstar$ population, 
tend to be more rapidly rotating, with disky isophotal shapes 
and central light cusps. The transition between the two occurs at approximately 
$M_{V}\sim-21.5$, and is commonly thought to derive from the difference between 
systems which have undergone just their initial, spheroid forming and gas-rich 
merger (which will dominate the less massive systems) and those which 
have undergone subsequent spheroid-spheroid dry mergers (which will 
dominate the most massive systems). 

Indeed, detailed numerical simulations 
have shown that gas-rich disk mergers, and only gas-rich mergers, 
reproduce the detailed distributions of kinematic properties of 
the less massive/rapidly rotating/disky/cuspy $\sim\lstar$ elliptical population 
\citep{cox:kinematics} -- including their rotation properties, 
kinematic misalignments, isophotal shapes, ellipticities, central light cusps 
\citep{mihos:cusps}, velocity profiles \citep{naab:gas}, 
kinematic subsystems \citep{hernquist:kinematic.subsystems},
and internal correlations \citep{robertson:fp}. 
Likewise, mergers of kinematically hot systems, i.e.\ spheroid-spheroid mergers, 
with little gas content, are required to produce the combination of 
boxy isophotal shapes, anisotropy, and low rotation seen in most massive ellipticals 
\citep{naab:dry.mergers}, and the commonly adopted theory of ``scouring'' by a binary black hole 
in the formation of central cores also 
requires that the mergers have very little cold gas content (since even 
$\sim1\%$ of the stellar mass in cold gas falling to the center is $\gg\mbh$, and 
would allow for rapid coalescence of a merging binary). 
Mergers of disks, even when gas-free, cannot reproduce the combination of 
low ellipticities and little rotation seen in the most massive spheroids 
\citep{cox:kinematics}.

\begin{figure}
    \centering
    \figexpand
    \plotter{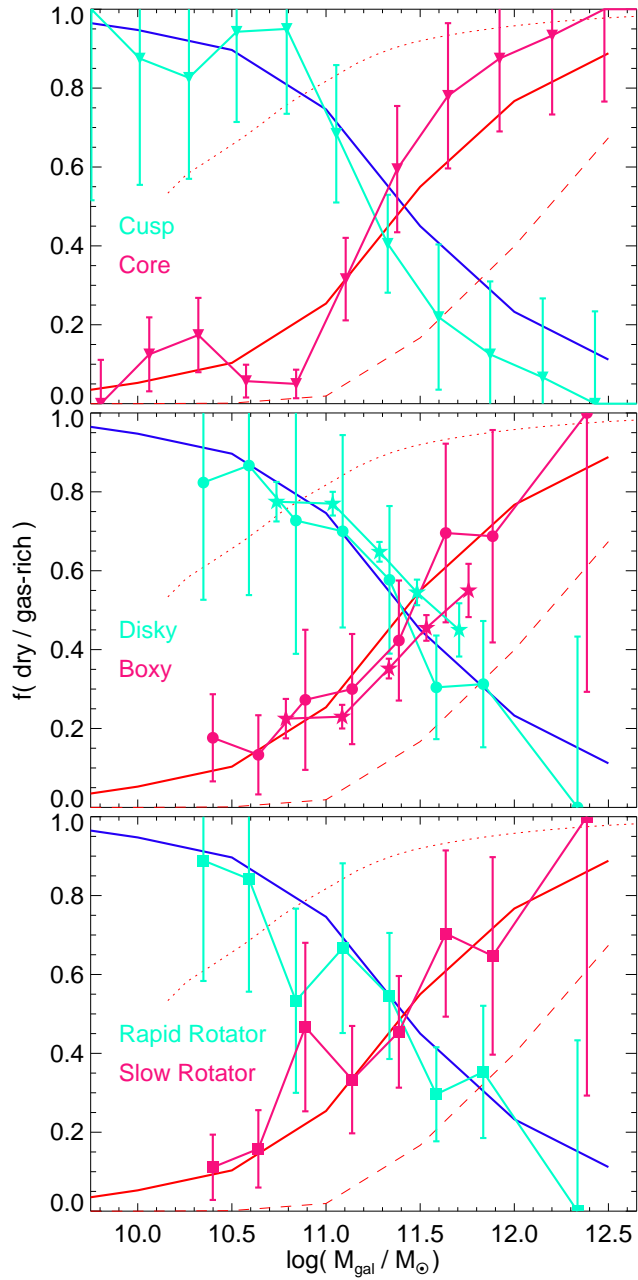}
    \caption{Predicted fraction of spheroids for which the last merger was a gas-rich, 
    spheroid-forming 
    major merger (blue), or for which 
    the last merger was a subsequent spheroid-spheroid (dry) 
    major merger (red), as a function of galaxy stellar mass. 
    We compare with the observed fraction of cusped or cored-central profile 
    ellipticals \citep[][]{lauer:bimodal.profiles}, the fraction of ellipticals with 
    disky ($a_{4}/a>0$) or boxy isophotal shapes \citep[][circles and 
    stars, respectively]{bender:ell.kinematics.a4,pasquali:boxy.frac}, 
    and the fraction of rapidly ($\log{(v/\sigma)^{\ast}}>-0.15$) or slowly rotating/isotropic  
    ellipticals \citep[][]{bender:ell.kinematics}. The dichotomy between 
    elliptical types is reproduced well, if dry mergers form 
    cored, boxy, slowly rotating remnants (as suggested by numerical simulations). 
    In each panel, the solid lines are the predictions of our merger model, 
    dotted and dashed lines show the predictions of 
    secular and halo quenching models, respectively (see Figure~\ref{fig:dry.models}; 
    for clarity just the dry merger fractions are shown). 
    \label{fig:dry.obs}}
\end{figure}

Figure~\ref{fig:dry.obs} therefore compares the fraction of 
cusp/core, disky/boxy, and rotating/isotropic ellipticals as a function of 
galaxy stellar mass to our estimate of the fraction of $z=0$ 
systems for which the last major merger was a 
gas-rich, spheroid-forming merger, or for which 
the last merger was a (subsequent) spheroid-spheroid 
dry merger. The agreement is good, for all three indicators. Both the trend in the 
fraction as a function of mass, and the transition at 
$\mgal\sim2-3\times10^{11}\,\msun$ are predicted by our model. 
This transition point is robust, with a rough $\sim0.2$\,dex 
systematic uncertainty owing to the exact version of our model which is used to calculate 
the merger histories (within the range of uncertainties from the observations). 

\begin{figure}
    \centering
    \figexpand
    \plotone{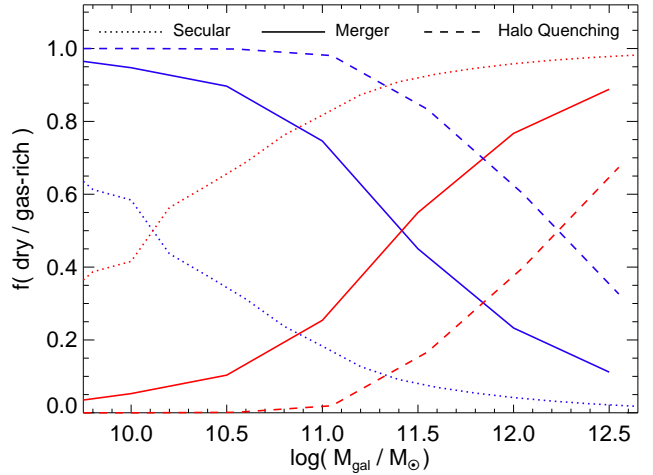}
    \caption{Predicted fraction of spheroids for which the last merger was a gas-rich, 
    spheroid-forming 
    major merger (blue), or for which 
    the last merger was a subsequent spheroid-spheroid (dry) 
    major merger (red), as a function of galaxy stellar mass as 
    in Figure~\ref{fig:dry.obs}, but for different models. The predictions of our 
    merger model ({\em solid}) are compared to those 
    from a pure secular ({\em dotted}) or pure halo quenching ({\em dashed}) 
    models. Secular models quench and form bulges via disk instability, so 
    most mergers even at low mass are dry/spheroid-spheroid; halo quenching models do not 
    prevent spheroids from re-forming disks below the halo quenching mass, 
    so only the most massive mergers are dry/spheroid-spheroid. Both predict 
    a transition point between cuspy/disky/rapidly rotating and 
    cored/boxy/slow rotating ellipticals an order of magnitude discrepant from 
    that shown in Figure~\ref{fig:dry.obs}.
    \label{fig:dry.models}}
\end{figure}

This additionally puts strong constraints on other models, 
shown in Figure~\ref{fig:dry.models}. In a pure secular model, 
most ellipticals are formed via disk instabilities -- this is already 
in conflict with the kinematic arguments in \S~\ref{sec:intro} (which 
find that disk instabilities generically form {\em pseudobulges}, not 
the classical bulges that dominate the spheroid population at the 
masses of interest here), but in addition, these systems 
therefore are already 
gas exhausted and quenched by the time they undergo their first major merger. 
Nearly {\em all} major mergers in such a model, then, constitute dry, spheroid-spheroid 
mergers. Adopting our calculated merger histories as a 
function of mass (which are not sensitive to our model for 
the colors and morphologies of the merging systems), but using the pure secular 
model criteria (as in Figure~\ref{fig:red.frac.bivar}) to determine whether the progenitor 
galaxies are already 
red (i.e.\ whether or not the merger is dry), we obtain the 
prediction that the transition to dominance of dry, spheroid-spheroid mergers 
should occur at masses an order of magnitude lower, $\mgal\sim10^{10}\,\msun$, 
an order-of-magnitude contradiction with the observations. 

Likewise, the simplest pure halo quenching models fail to reproduce the observed 
elliptical dichotomy. In such models, a substantial fraction of galaxies will 
experience their first major merger before the system crosses the quenching 
halo mass threshold. Since they are below this threshold, the system will 
re-accrete gas and re-form a disk, even for major mergers occurring 
just a short time ($\sim$\,a couple Gyr) before the halo grows sufficiently 
massive to cross this threshold. The next major merger will therefore 
be assumed (in the model) to be a 
gas-rich disk merger, instead of a dry spheroid-spheroid merger. 

We again calculate the 
effects of this in Figure~\ref{fig:dry.models}, using our merger histories but 
assuming that all systems below the halo quenching mass threshold of 
\citet{croton:sam} (at each redshift) will re-accrete and remain gas-rich. 
The result is that only the most extremely massive systems, which 
crossed the halo quenching mass threshold at early times, 
have had sufficient time to then undergo subsequent multiple mergers 
(yielding at least one spheroid-spheroid dry merger). We obtain the 
prediction that the transition point between 
elliptical types should occur at masses an order of magnitude higher 
than in our merger-driven model, at $\mgal>10^{12}\,\msun$, 
once again an order-of-magnitude contradiction with the observations. 

This is further demonstrated in the recent analysis by \citet{kang:boxy.frac.sam}, 
who consider the predictions for the number of boxy ellipticals 
in a similar halo quenching model.  For the reasons given
above, their model predicts that a negligible ($\ll 10\%$) fraction of 
early-type galaxies have undergone a true dry (spheroid-spheroid) 
merger, at all masses $\mgal\lesssim10^{12}\,\msun$. In order to 
match the observations in Figure~\ref{fig:dry.obs}, they are forced 
to assume that {\em any} major merger with a gas fraction 
$f_{\rm gas}<0.1$ produces a boxy elliptical. As a consequence, 
such a model
predicts that $\sim1/3-1/2$ of ``boxy'' systems are actually formed 
in mergers of two disk galaxies (low gas-fraction, 
Milky Way-like disks), with the remaining $\sim1/2-2/3$ 
formed in what we would call gas-rich, mixed morphology mergers. 
However, numerical simulations of major mergers of disk-dominated 
galaxies with gas fractions $f_{\rm gas}\lesssim0.1$, and 
kinematic analysis of comparable local merger remnants \citep{rothberg.joseph:kinematics},
have clearly established that such mergers {\em do not}, in fact, 
generically produce boxy ellipticals. They instead produce systems 
resembling disky ellipticals \citep{naab:gas}, with substantial 
central cusps \citep[the cusps do not disappear even at low $f_{\rm gas}$;][]{mihos:cusps}, 
and high ellipticities \citep{cox:kinematics}. 

If we instead begin knowing the properties of mergers that form 
different types of ellipticals, the observations lead us to conclude that some form 
of quenching must be able to operate, at least temporarily, in massive spheroids 
after their first formation epoch, in order that they be truly dry/spheroidal in 
subsequent mergers. A similar conclusion is reached by \citet{naab:dry.mergers}, 
who find, using numerical simulations to measure the distribution of 
spheroid isotropy/anisotropy that would be observed in a remnant of a merger of 
specific types of progenitor galaxies, that matching the trend and transition to the 
dominance of anisotropic galaxies requires the quenching of all systems with massive 
bulges ($\gtrsim3\times10^{10}\,\msun$, in their case), effectively identical to 
our merger quenching criterion.
This demonstrates the strong constraints that can be placed on models for 
how systems quench and become red galaxies, given the 
specific kinds of galaxy mergers required to produce the correct distribution of 
elliptical kinematic properties as a function of mass. A more detailed 
investigation combining these cosmological predictions with 
detailed numerical simulations, to study the effect of such mergers 
on the kinematics, internal correlations, and redshift evolution of 
massive ellipticals is outside the scope of this paper, but is an important
subject of future work.

\section{The Physics of Quenching}
\label{sec:quenching}

We now turn to a discussion of the physical mechanisms by which mergers 
might both terminate significant star formation, and 
result in a system which can maintain relatively low star formation rates. 
It is not our intent in this discussion to prove that a particular 
mode of feedback, for example, {\em must} quench subsequent star formation, 
but rather to highlight the physical processes that operate in mergers 
and their possible or likely effects on the intergalactic medium (IGM)
and subsequent cooling of halo gas.

To do so, we examine a large suite of hydrodynamical simulations of 
disk galaxies and major mergers between them, described in 
detail in \citet{robertson:fp}.
The simulations are high-resolution 
(spatial resolution $\sim20\,{\rm pc}$ in the best cases), 
fully hydrodynamic calculations
which incorporate a self-consistent, observationally motivated model 
for a multiphase interstellar medium,
star formation, supernova feedback, and black 
hole accretion and feedback \citep[for details, see][]{springel:multiphase,springel:models}. 
We construct stable, equilibrium 
disk galaxies to either merge or evolve in isolation as described in 
\citet{robertson:fp}, with a dark matter halo, gas and stellar disk, and bulge 
component relevant for observed galaxies of the given mass and 
redshift, with e.g.\ the 
scale length of these components set by the appropriate concentration 
and spin parameter as a function of mass and redshift. We specifically consider 
a subset of disks with baryonic masses 
$\mgal \approx 10^{10},\,10^{11},\,10^{12}\,\msun$, and initial simulation gas 
fractions of $f_{\rm gas}=0.4$ and $0.2$. In our merger simulations, we 
place two identical disks with a relative inclination of $\sim60^{\circ}$ (representative 
of most random encounters) on a parabolic orbit and allow the system to evolve 
until it has completely relaxed (usually $\sim2.5-3$\,Gyr after the merger).
The simulations were performed using the code
Gadget-2 \citep{springel:gadget}, a fully conservative \citep{springel:entropy}
implementation of smoothed particle hydrodynamics (SPH).

We show just the results from these cases in what follows, for simplicity, but 
note that we have surveyed
a much wider parameter space in 
\citet{robertson:fp,hopkins:qso.all,cox:kinematics}, 
varying masses from $\mgal \sim 10^{8} - 10^{13}\,\msun$, 
gas fractions $\fgas = 0.05 - 1$, concentrations, bulge-to-disk ratios, 
and (in mergers) orbital parameters, relative disk 
inclinations, and merger mass ratios. We ultimately find qualitatively similar results 
in all these cases, and for our purposes the subset of simulations shown is 
representative of the important qualitative effects. 

In each simulation, we assume any stars 
present at the beginning were formed according to the 
best-fit observed $\tau$-model star formation history for disks of the given mass 
in \citet{BelldeJong:disk.sfh} (appropriate for the redshift at which the 
simulation is initialized). The stars formed during the simulation 
have ages and metallicities determined self-consistently. Knowing 
the star formation and enrichment history of all stars in the simulation, we 
integrate to calculate the mean $(B-V)$ color of the galaxies at 
each time, using the stellar population synthesis models of 
\citet{BC03} with an assumed \citet{chabrier:imf} IMF 
(similar in the predicted colors to our generally adopted diet Salpeter IMF). 
Because the simulation also includes gas, we can self-consistently 
integrate along the line of sight to all star particles and calculate 
the appropriate dust reddening and extinction, 
following \citet{hopkins:lifetimes.methods}. However, because we 
are primarily interested in times 
after the merger (i.e.\ after most gas is exhausted in star formation) 
and average trends, we find this makes little difference.

\subsection{``Transition'': Termination of Star Formation in Major Mergers}
\label{sec:quenching:transition}

\begin{figure*}
    \centering
    \figexpand
    \plotone{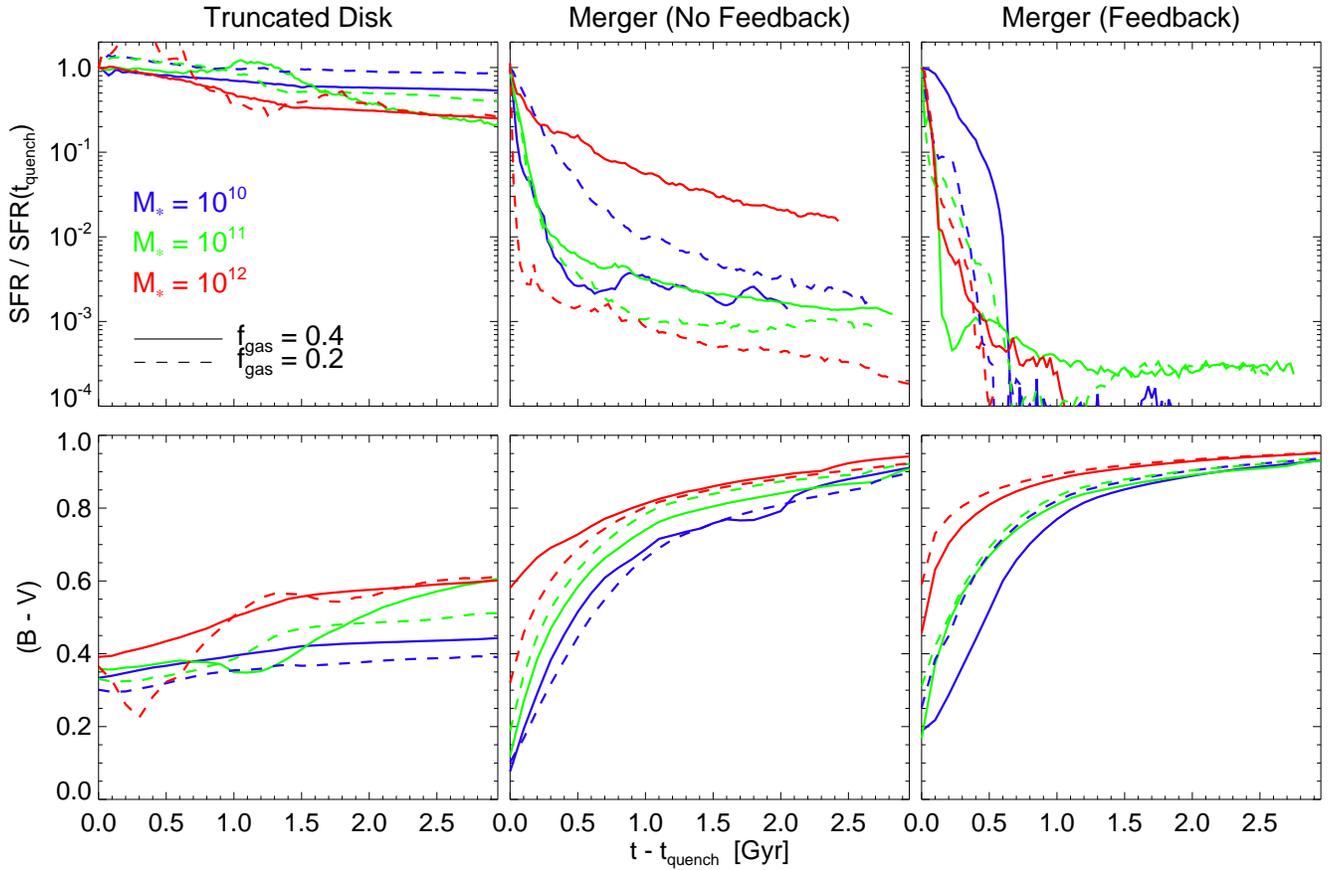}
    \caption{Evolution of star formation rate (relative to that at $t=0$) 
    and $(B-V)$ optical colors of galaxies of different initial 
    baryonic masses and gas fractions (as labeled), 
    in high-resolution hydrodynamic simulations. {\em Left:} Evolution of an isolated 
    quenched or ``truncated'' disk which is completely cut off from 
    accretion/external gas supplies at $t=0$. {\em Center:} Evolution of a 
    merger remnant after the final galaxy coalescence/starburst phase at $t=0$. 
    No feedback is included (i.e.\ the decrease in star formation rate derives 
    entirely from gas exhaustion and shock-heating). {\em Right:} Evolution of a 
    merger remnant, with feedback in the form of starburst-driven 
    and quasar-driven winds. The mergers rapidly redden to 
    the red sequence ($(B-V)\gtrsim0.8$) in $\lesssim1\,$Gyr, but 
    an isolated disk (even with secular instabilities operating) remains blue. 
    \label{fig:quench.colors}}
\end{figure*}

It is relatively easy to see how a major merger can terminate star formation 
in an immediate sense. The rapid consumption of gas in the final stages of 
the merger, potentially coupled with expulsion by feedback mechanisms, 
allows for a sharp truncation in star formation. Figure~\ref{fig:quench.colors} 
illustrates this. 

We first consider 
simulations of ``truncated'' disks; i.e.\ disks which are completely cut off 
from a gas accretion supply. We construct appropriate disks of the 
masses and gas fractions shown in Figure~\ref{fig:quench.colors} 
and evolve them in isolation (allowing no further gas accretion). 
Technically, in terms of the stellar populations and disk properties, 
the plot assumes that disk accretion is truncated at $z=2$, appropriate 
for most of the star formation in present-day early type galaxies 
\citep[e.g.][and references therein]{gallazzi06:ages}, but the qualitative result is 
almost identical regardless of when we initialize the simulation. 
The star formation rate, plotted as a fraction of that
at the onset of the simulation (since the optical colors here 
are primarily influenced 
by the relative decline in star formation), decays weakly. In 
fact, this drop is similar to that expected in the simplest models. 
For any disk 
which obeys a $\tau$-model star formation history 
$\dot{M} \propto \exp{(-t/\tau)}$ prior to truncation, and a 
Kennicutt-Schmidt star formation law $\Sigma_{\rm SF} \propto \Sigma_{\rm gas}^{1.4}$
\citep{kennicutt98}, it is straightforward to calculate the 
subsequent evolution in the star formation rate if the disk accretion 
is instantaneously truncated at a time $t_{f}$ (since the disk 
size and baryonic mass should no longer evolve) -- 
it will evolve as $\dot{M}\propto (1+ [t-t_{f}]/t_{0})^{-7/2}$, 
where $t_{0}\approx\tau$ is a constant timescale which depends in detail 
on the gas fraction, gas mass profile, and time of 
star formation truncation ($t_{0}=0.72\,\tau$ for a 
$10^{11}\,\msun$ exponential disk with $\fgas=0.4$ truncated at $z=2$). 

Figure~\ref{fig:quench.colors} next shows the star formation rate and colors 
of merger remnants, after the merger itself. We consider two cases -- first, 
with no feedback (i.e.\ no stellar winds, and no black hole accretion or feedback) 
included, and second, with a standard, observationally calibrated and 
relatively mild prescription for starburst-driven winds \citep[with a 
wind outflow rate roughly half the star formation rate; see][]{cox:winds} and BH accretion 
and feedback \citep[such that the BHs self-regulate at masses appropriate for 
the observed $\mbh-\sigma$ relation; see][]{dimatteo:msigma}. Star formation 
rapidly falls by orders of magnitude after the merger, even in the ``no feedback'' case, 
as the majority of the gas supply has been rapidly consumed in a central 
starburst and much of the remaining gas shock-heated into an X-ray halo 
\citep{cox:xray.gas}. With feedback, the suppression is even more complete, as 
stellar and quasar-driven winds clean up the last remaining traces of 
star-forming gas. 
In both cases, the remnants redden extremely rapidly, requiring 
less than one Gyr to reach the red sequence. We do caution that, in mergers of 
extremely gas-rich disks, feedback may be necessary to redden so 
rapidly -- \citet{springel:red.galaxies} showed this for $100\%$ gas disks; but in any 
case the level of feedback required is reasonable (comparable to that used here), 
and this is probably only relevant for the highest-redshift mergers. 

Two conclusions emerge from Figure~\ref{fig:quench.colors}. 
In the case of a 
truncated disk, the decline in star formation rate is gradual, so the $(B-V)$ 
colors redden very slowly. Even in a truncated $10^{12}\,\msun$ disk 
after 3\,Gyr, the galaxy colors are significantly bluer ($(B-V)\sim0.6$) 
than those of a typical red sequence galaxy ($(B-V)>0.8$). Furthermore, 
although our simulations allow for disk instabilities (see Figure 6 of
Springel et al. 2005b),
and do form 
spiral structure and even bars (seen in the 
small variations in star formation rate), this (even in the most massive, gas rich cases) 
does not consume sufficient gas to quench the disk. It is extremely difficult for 
secular mechanisms to exhaust {\em all} the gas, especially in the outer, low density 
regions of disks, and only a small continued rate of star formation is 
necessary to keep the galaxies blue. 

Furthermore, it is observed 
locally and at redshifts up to $z\sim1$ \citep{bell:combo17.lfs} that only a 
small fraction of galaxies occupy the ``green valley'' between blue cloud and red 
sequence. Assuming that $\sim1/2$ of $\sim\mstar$ galaxies must cross the 
green valley since $z=1$ \citep[roughly what is expected from comparison of the 
mass/luminosity functions, e.g.][]{martin:mass.flux}, a slow reddening such as that of the truncated 
disk in Figure~\ref{fig:quench.colors} would imply as many as $\sim1/4-1/3$ of 
all $\sim\mstar$ galaxies should occupy this region, compared to the $\ll10\%$ 
observed \citep{bell:combo17.lfs}. Simply put, this would eliminate or 
completely smooth out the observed 
strongly bimodal color-magnitude distribution \citep[e.g.][]{baldry:bimodality}, 
at least in moderately massive galaxies (our simulations are obviously not 
meant to be applied to e.g.\ dwarf satellites). 
In contrast, even gas-rich merger remnants with no feedback redden rapidly to 
the red sequence, with a timescale for reddening 
of $\lesssim1\,$Gyr that is completely consistent with the observed 
color bimodality and small fraction of galaxies in the ``green valley.''

\subsection{``Maintenance'': How Is Later Star Formation Suppressed?}
\label{sec:quenching:maintenance}

It is apparent that merger remnants redden rapidly onto the 
red sequence.  However, whether or not they 
can stay on the red sequence for significant periods of time is
less certain. 
In other words, although mergers 
easily {\em terminate} star formation, they will not remain 
as long-lived red galaxies unless they also {\em maintain} low 
levels of accretion and star formation.

\subsubsection{Is there a need to do so? The ``No Feedback'' Solution}
\label{sec:quenching:maintenance:nofeedback}

One possibility is that this maintenance is trivial. Roughly half the 
present mass density in red galaxies is built up since $z=1$, and the typical 
host halo of a $\sim\mstar$ red galaxy at $z=0.5$ will grow only by $\sim0.2$\,dex 
to $z=0$. While this is still enough fractional growth ($\sim50\%$) to make the galaxy 
blue, if all the newly accreted gas were to cool immediately and form new stars, 
it is unlikely that this small amount of gas at the virial radius could cool and 
infall within the $\sim5\,$Gyr timescale to $z=0$. Indeed, the 
``cooling flow problem'' appears to be a problem for only the most massive 
clusters at low redshifts 
\citep[e.g.][but see also \citet{chen:cooling.flow.small}]{best:radio.loudness,vikhlinin:low.cooling.flows.at.highz}, 
which suggests that cooling flows 
are, in general, a late-forming phenomenon just now becoming relevant, 
and perhaps were never suppressed in the past. Furthermore, many 
``central'' galaxies do not actually reside at the exact center of their group or 
cluster potential \citep{mulchaey:midz.groups,jeltema:midz.groups}, 
as is naively assumed in most analytic models, 
which makes the formation of cooling flows less efficient. 
Recent high-resolution simulations \citep{naab:etg.formation,keres:prep} 
do suggest that, without any AGN or stellar feedback, the combination of virial shocks, 
compression, and kinematic heating by clumpy accretion flows 
can prevent substantial cooling at $z\lesssim1$, and that simple gas exhaustion 
via star formation can quickly eliminate most of the low cooling-time gas. 

However, this is not entirely satisfactory, at least in the simplest sense. 
First, cooling flows still do appear to be a problem in these systems -- and 
the most massive galaxies are almost uniformly red \citep{baldry06:redfrac.vs.m.env}, 
they do not appear to 
be recently accreting/star-forming or ``becoming'' blue
\citep[but see][who reach the opposite conclusion for 
BCGs with large cooling flows]{rafferty:cooling.flows.are.growing}. Second, 
even in the moderate-mass halo case considered above, the free-fall 
time of the gas accreted since $z=0.5$, $\sim2\,$Gyr, is sufficiently 
short that a cooling flow problem remains a possibility. The problem 
also becomes more severe at high redshifts, where cooling 
rates can be a factor $\sim100$ higher 
than at $z=0$ (scaling $\propto n \propto (1+z)^{3}$). 
Finally, essentially all implementations of galaxy formation 
models which attempt to account for gas accretion and cooling with 
various prescriptions have found that feedback of some 
kind is necessary to prevent new accretion in massive galaxies 
\citep[e.g.][]{birnboim:mquench,binney:cyclic.feedback,granato:sam,
scannapieco:sam,keres:hot.halos,
monaco:feedback,croton:sam,dekelbirnboim:mquench,cattaneo:sam}.

\subsubsection{Can Quasar/Starburst Feedback Completely Suppress Future Cooling?}
\label{sec:quenching:maintenance:nuke}

It is therefore natural to examine the feedback effects involved in (or stemming 
from) major mergers. We identify four primary feedback mechanisms: 

{\bf (1) ``Kinematic'' Feedback:} Mergers themselves stir large 
quantities of gas, allowing relatively hot and cold gas from the inner and outer 
regions of a shared halo to mix, and generally increasing the 
cooling time 
significantly and disrupting any cooling flows ongoing or in formation. This is 
seen both in simulations \citep{naab:etg.formation,keres:prep,cox:xray.gas} and X-ray observations of 
galaxy groups \citep{jeltema:midz.groups,vikhlinin:low.cooling.flows.at.highz}. 
Furthermore, tidal shocks in the merger itself 
heat a significant quantity of gas to temperatures well above those that can 
efficiently cool in a Hubble time (the reason for the low star formation rates 
at late times in Figure~\ref{fig:quench.colors}, even in the ``no feedback'' case). 

{\bf (2) Starburst-Driven Winds:} It is known from local 
measurements \citep[e.g.][]{kennicutt98} 
and also suggested in high-redshift studies \citep{erb:lbg.metallicity-winds} that a high surface 
density of star formation inevitably results in strong galactic winds. Presumably 
driven by a combination of young stellar winds and supernovae, the energy 
coupling efficiency appears to be high (order unity), and simulations 
demonstrate that the observations are well-reproduced for reasonable, theoretically 
expected mass-loading efficiencies \citep[$\eta\sim0.5$, where 
$\dot{M}_{\rm wind} = \eta\,\dot{M}_{\ast}$, with possible mass dependence from a 
momentum-based escape approximation; see][]{cox:winds,oppenheimer:outflow.enrichment}. 
These will act throughout a merger, 
and are a powerful integrated source of feedback, although not as impulsive 
as quasar-driven outflows \citep[e.g.][]{lidz:proximity}.

{\bf (3) ``Quasar'' Feedback:} Quasars are known to often exhibit 
strong outflows \citep[for a review, see][]{veilleux:winds} 
and to have a large effect on the ionization and 
temperature state of the inner regions of their host galaxies 
\citep[e.g.][and references therein]{laor:warm.absorber,krongold:seyfert.outflow,rupke:outflows}. 
At the brightest luminosities, a large fraction ($\sim40\%$) of sightlines 
to quasars see highly energetic ($\gtrsim$\,several $10^{3}\,{\rm km\,s^{-1}}$) 
broad absorption line (BAL) outflows, and it is likely that all 
bright quasars exhibit some such BALs \citep[although they may not be visible owing 
to geometric effects;][]{reichard:bal.properties,elvis:outflow.model,
gallagher:bal.xray,gallagher:bal.submm}. 
These feedback mechanisms {\em must} be able to have a dynamical 
effect in some sense on the host, in order to suppress accretion onto the 
central BH once it reaches the limit of the $\mbh-\sigma$ relation. Recall, 
if only $\sim0.1\%$ of the initial galaxy gas mass were to survive a merger 
and make its way to the center of the galaxy, it would (without BH feedback) 
drive the central BH off the $\mbh-\sigma$ relation by more than 
the observed scatter (while having almost no effect on $\sigma$). 

What is 
less certain is the effect such feedback has on the largest galactic 
scales. In most models, it is inevitable that the small-scale 
wind, heating, or pressurization required to halt accretion and produce the 
$\mbh-\sigma$ relation will indeed generate a galactic outflow, and 
recent high-resolution, self-consistent simulations of quasar feedback 
and accretion disk winds 
imply the formation of powerful kinematic outflows that will couple on 
larger scales \citep{proga:disk.winds}. Indeed, an increasing number of 
bright quasar hosts have now been observed in which jets or winds 
appear to be strongly impacting the host galaxy and halo gas 
\citep{zirm:radio.gal.feedback,nesvadba:radio.gal.feedback,
nesvadba:smg.feedback,reuland:quasar.feedback}, or in 
which BAL quasar 
winds are entraining gas at $\sim10^{3}-10^{4}\,{\rm km\,s^{-1}}$ 
velocities on $\sim$\,kpc scales \citep{dekool:large.outflow.1,dekool:large.outflow.2,
gabel:large.outflow}. Indeed, high velocity winds at (or beyond) galactic scales 
appear to be ubiquitous in
post-starburst galaxies at moderate redshifts, and trace a continuum in outflow 
properties with bright quasars \citep[typically with low-luminosity AGN consistent 
with fading from a recent peak of starburst and quasar activity;][]{tremonti:in.prep,ganguly:qso.outflows}.
The integrated energy in this feedback is, from simple energetic arguments, comparable 
to that from stellar winds \citep{lidz:proximity}.
However, the timescale is much shorter -- 
the BH gains most of its mass (releases most of its energy) in less than a 
Salpeter time $\sim10^{7.5}\,$yr, whereas most of the stars are typically formed 
over a timescale $\gtrsim10^{9}\,$yr. Therefore, even in the most conservative 
models, the {\em power} in such quasar-driven winds is $\sim10-100$ times 
greater than that in typical starburst-driven winds.

As a result of the short timescales associated with this process, and
because the energy or momentum is injected on scales small compared with
those of entire galaxies, the impact of quasar feedback is explosive in
nature.  Indeed, \citet{hopkins:seyferts} and \citet{hopkins:faint.slope}
have demonstrated that the outflows in the simulations caused by this
phenomenon are well-approximated by a generalized blast-wave solution.

{\bf (4) ``Radio-Mode'' Feedback:} As coined by \citet{croton:sam}, 
this refers to a maintenance mode of 
feedback, including e.g.\ the inflation of radio bubbles in clusters at 
relatively low accretion rates \citep[e.g.][and references therein]{fabian:perseus,dunn:bubble.heating,
allen:jet.bondi.power,sanders:perseus}, but also the driving of weak winds from 
radiatively inefficient accretion flows \citep[e.g.][]{NY94} and X-ray heating of 
nearby gas. Essentially, this is a blanket term for all feedback mechanisms 
which depend on a massive BH at relatively low levels of activity 
(low Eddington ratios), in which most massive BHs spend most of their 
lifetimes (typically since $z\sim2$). It explicitly does {\em not} include ``quasar-mode'' 
feedback, representative of the high-Eddington ratio, high-power output effects 
described above (and it also does not exclude radio jets as a feedback 
mechanism in the ``quasar-mode''). 

This mode of feedback operates over long timescales ($\sim\tH$) 
and would not occur under the high-Eddington ratio conditions of mergers. 
Nevertheless, we include it here because it is linked to mergers in a critical way: 
almost universally, the forms of ``maintenance'' feedback 
require the presence of a relatively massive BH \citep[e.g.][]{dekelbirnboim:mquench,
binney:cyclic.feedback,croton:sam,sijacki:radio}. A massive 
BH empirically 
requires a massive spheroid, which requires a major merger. In other words, 
{\em there is no ``radio-mode'' feedback without major mergers.}
Indeed, a comparison of the local mass density of supermassive black
holes with the luminosity density of quasars integrated over redshift
\citep{soltan82,hopkins:bol.qlf} indicates that the mass growth
through the radio mode must be negligible \citep{hopkins:old.age}, in
agreement with preliminary results from cosmological simulations of
these processes \citep{sijacki:radio,dimatteo:cosmo.bhs}.

Although the distinctions between these modes of feedback, and ultimately 
the detailed identification of the drivers of each are of great importance, these 
are questions outside the scope of this paper. A detailed comparison, for example,
of the effects of different modes of feedback on the IGM and their observable 
signatures will be the topic of a future paper (Hopkins et al., in preparation). 
For now, we simply wish to examine whether a reasonable integrated effect of 
such feedback could be to completely suppress future star formation in 
merger remnants. To do so, we return to the numerical simulations described above. 

\begin{figure}
    \centering
    \figexpand
    \plotone{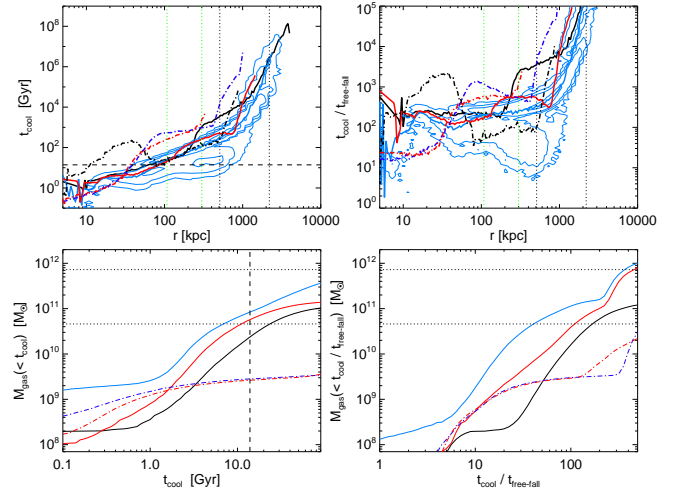}
    \caption{{\em Top:} Cooling time as a function of radius for typical relaxed 
    SPH simulation merger 
    remnants embedded in massive gaseous halos. We show both the absolute 
    value of the cooling time ({\em left}) and the cooling time relative to the 
    local free-fall time ({\em right}). Solid (dot-dashed)
    lines show the mass-weighted mean value at each $r$ for 
    galaxies with stellar masses $\sim10^{12}\,\msun$ ($\sim3\times10^{10}\,\msun$). 
    Different colors correspond to different initial halo gas profiles: 
    black assumes a relatively low total halo gas mass, red and blue a factor of several 
    higher gas mass (with a pre-merger isothermal or pressure-supported temperature profile, 
    respectively). 
    For one case we show 
    contours of the full gas distribution at $50,\ 5,\ 1$ and $0.1\%$ -- most of the gas is 
    close to the mean value. Vertical dotted lines show 
    the virial radii of the halos of both masses at $z=0$ (black) and $z=2$ (green).
    {\em Lower:} The integrated gas mass below a given cooling time ({\em left}) 
    or cooling time relative to free-fall time ({\em right}). Dotted horizontal lines 
    show the two galaxy stellar masses represented here. 
    Dashed line in left panels shows the Hubble time. 
    Feedback from a major merger heats a large quantity of gas and establishes a 
    hot or ``quasi-static'' halo.
    \label{fig:hot.halos}}
\end{figure}

First, we consider several merger simulations similar to those in Figure~\ref{fig:quench.colors}, 
with reasonable, observationally constrained feedback prescriptions from 
both star formation and quasars. We embed the progenitor systems in large gaseous halos 
meant to represent the gas both in the host halo and surrounding it 
(which will be accreted in the future). Specifically, we consider a range of gas mass 
for the halo, from the total gas mass expected within $R_{\rm vir}$ to several 
times this quantity (representative of that which will be accreted in the next few Gyr). 
We consider both an NFW profile for the gas and a uniform density distribution,
and calculate 
the initial gas temperature either assuming a uniform heating to the virial temperature or 
initial hydrostatic equilibrium. The halos are initialized appropriate for those at 
low redshifts, and the galaxies are otherwise constructed identically 
to those described in \S~\ref{sec:quenching:transition}. 
We evolve our merger simulations until they are 
relaxed (typically $\sim2-3\,$Gyr, as before), 
and calculate the cooling time \citep[including metal-line cooling, following][]{cox:xray.gas} 
for all remaining gas particles at the end of the simulation. 

Figure~\ref{fig:hot.halos} shows the results for several representative simulations. 
For clarity, we do not show the results of every simulation, but note that 
the qualitative behavior is, in all cases of a given mass, quite similar, with 
properties such 
as the initial gas density profile affecting only the details of the final gas profiles 
(not their general behavior as a function of mass and/or radius). We plot the 
gas cooling time as a function of radius -- by this time, the gas has relaxed and 
there is a reasonably well-defined cooling radius inside which the gas will 
cool in $\ll \tH$. The actual mass contained therein is not negligible -- 
only $\sim5-10\%$ can be added to the galaxy mass in a Hubble time. 
This is, in principle, sufficient to make the galaxy blue once again, however 
most of the cooling would happen at late times -- where another small burst 
of feedback could re-heat the gas and prevent this scenario. For galaxies 
moving to the red sequence at $z\lesssim1$ ($\sim1/2$ of present red galaxies), 
the suppression is even stronger -- only $\lesssim1\%$ of the post-merger galaxy 
stellar mass can cool by $z=0$, which is sufficiently small to 
ensure that the galaxy remains ``red and dead.'' 

These simulations, however, neglect the dynamical nature of accretion onto 
the dark matter halo with cosmic time, and do not include the 
{\em very} large relative gas mass accreted onto the most massive, early-forming 
systems. For example, a $\gtrsim{\rm few}\times10^{8}\,\msun$ BH forming at $z=2$ 
will live in a $\gtrsim10^{13}\,h^{-1}\,\msun$ halo, which will typically grow by 
more than an order of magnitude in mass to $z=0$.  In our complete SPH simulations, 
surveying this parameter space requires large boxes, external reservoirs of gas,
inclusion of cosmological effects, and long runtimes ($\sim\tH$), and is ultimately 
outside the scope of this paper. However, we can make some rough estimates 
of the qualitative effects from simple scaling arguments. 

Consider the feedback energy which couples to the galactic ISM during a 
merger ($E_{\rm merger}$). The integrated 
feedback energy injected by the ``quasar mode'' over the course of the merger will be 
(given that most of the BH mass is gained in this phase) approximately 
$E = \eta\,\epsilon_{r}\,\mbh\,c^{2}$, where $\epsilon_{r}\approx0.1$ is the 
radiative efficiency and $\eta$ is the feedback coupling efficiency ($\eta\sim0.05$ 
in our simulations in order to yield the appropriate normalization of the 
$\mbh-\sigma$ relation). As noted above, the total feedback energy 
from star formation is of a comparable order (although 
it operates over a much larger timescale, so only some fraction 
will couple during the merger itself), so we can subsume it into this 
scaling (since $\eta$ is uncertain anyways, it can effectively include the 
maximal factor $\sim2$ addition from stellar winds and supernovae). 
The feedback from BH growth and at least the final, peak 
starburst phase will couple in a short time, $\sim10^{7.5}$ years (the 
timescale for the final $e$-foldings of BH growth), much shorter than the 
dynamical time in the outer regions of the halo. 

Assuming, therefore, that the ``merger feedback'' 
(by which we mean the combined feedback from quasar, starburst, and 
kinematic effects -- although the latter are energetically sub-dominant) 
creates a strong shock (true in nearly all of our 
simulations), the post-shock temperature inside the virial radius of the 
halo will be approximately 
$T_{\rm shock} \approx \alpha\,c\,T_{\rm vir}$, where $c$ is the halo 
concentration and $\alpha$ is a coefficient of order unity which depends in detail on the 
halo gas profile, baryon fraction, and metallicity \citep[we follow][who adopt 
standard values for these quantities, and obtain $\alpha\approx0.5$]{dekelbirnboim:mquench}. 
The resulting cooling time of the shocked gas near the virial radius is 
then (for the same parameters) roughly
$t_{\rm cool}\approx 8.3\,\Delta_{200}^{-1}\,(1+z)^{-3}\,(T_{\rm shock}/10^{6}\,{\rm K})^{2} \, \tH$, 
where $\Delta_{200}\sim1$ is 
the virial overdensity at the given redshift relative to a value of $200$ 
\citep[and we approximate the cooling function around the 
temperatures of interest following][]{sutherlanddopita93}. (Of course, 
most of the gas relevant for cooling will be at smaller radii and $\Delta_{200}\gg1$, 
but we simply wish to illustrate the relevant scalings.) 

If we were to rely on one feedback event alone to suppress all cooling until 
$z=0$, we would require two basic criteria. 
First, this clearly requires that $T_{\rm shock}$ be sufficiently 
high such that $t_{\rm cool} > \tH(z)$, i.e.\ $T_{\rm shock} > T_{\rm crit,\,H}
\approx 4\times10^{5}\,{\rm K}\,[\Delta_{200}\,f_{H}\,(1+z)^{3}]^{1/2}$, 
where $f_{H}$ is the fractional lookback time to redshift $z$. 
Second, the coupled feedback energy must be sufficient to heat {\em all} of the 
total $z=0$ halo gas content to these temperatures -- i.e.\ the 
total mass which can be shocked, 
$M_{\rm shock} = \mu\,m_{p}\,E_{\rm merger} / (3/2\, k\,T_{\rm crit,\,H})$ 
(where $\mu=0.59$ for pristine gas) must be equal to or greater 
than the total gas mass which will be accreted by $z=0$ and therefore 
which must be prevented from cooling. 
The first criterion is satisfied for all moderate halo masses of interest 
(although it may be that low-mass halos at high redshifts 
$z\gtrsim2$ have difficulty shocking to sufficiently high temperatures), 
and this is borne out by direct comparison with the post-shock temperatures 
in our simulations at all redshifts. It is at least likely that some of 
the surrounding gas will be shocked to very high temperatures -- the more 
interesting question is how this mass compares to the total mass 
that will be accreted and (potentially) otherwise cool by $z=0$. 

\begin{figure}
    \centering
    \figexpand
    \plotone{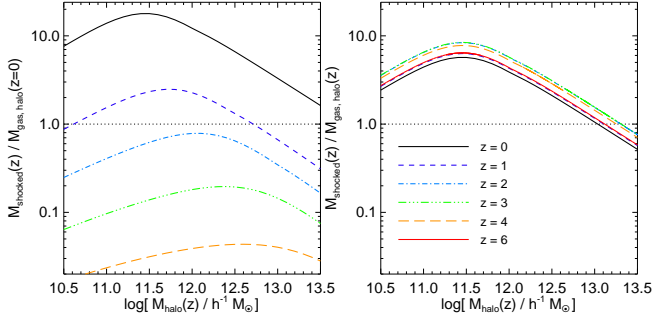}
    \caption{{\em Left:} Total gas mass 
    (relative to the total gas mass of the halo at $z=0$) 
    which can be shocked by a merger-induced 
    feedback-driven outflow/blastwave at the given redshift and host halo mass 
    above temperatures for which 
    the cooling time becomes longer than the Hubble time (i.e.\ total 
    fraction of the $z=0$ halo baryon content for which cooling can be completely 
    suppressed until $z=0$ by a typical merger at the given redshift). 
    {\em Right:} Total gas mass (relative to the total 
    gas mass of the halo at the given redshift) which 
    can be shocked to the critical shock stability temperature 
    (following \citet{keres:hot.halos,dekelbirnboim:mquench}) 
    above which the cooling time is longer than the (instantaneous local) 
    gas compression or free-fall time and a quasi-static halo is established. 
    Feedback from a major merger alone can quench all 
    future accretion in halos below the traditional ``hot halo'' mass threshold 
    at moderate redshifts ($z\lesssim2$), and can easily 
    establish a ``hot halo'' within the virial radius at all 
    redshifts.
    \label{fig:blastwaves}}
\end{figure}

Given our expectation for the average galaxy, and corresponding BH mass, in a 
halo of a given mass at some redshift, 
Figure~\ref{fig:blastwaves} compares the mass 
that can be shocked (given the energetic criterion above) to that accreted by $z=0$. 
In all cases, the feedback 
is able to shock-heat up to several times the initial galaxy mass, and 
we crudely expect the shock to propagate to several times the initial virial radius of the galaxy. 
However, the implications of this can be quite different for 
halos of different masses. Low mass halos, even at $z=2$, grow by a relatively small amount. 
For example, an average $10^{11}\,h^{-1}\,\msun$ halo at $z=2$ grows by a 
factor $\sim5$ to $z=0$ (so the feedback from the merger need only 
shock several times the galaxy mass in external gas to prevent all 
future accretion), but an average $10^{13}\,h^{-1}\,\msun$ halo at $z=2$ grows 
by a factor $\sim25$ to $z=0$. In small halos, then, feedback from a merger, 
at least at redshifts $z\lesssim2$, may be able to completely prevent future 
accretion, without the need to invoke any maintenance mode of feedback. 
In large halos, however, there is too much continued accretion 
and growth at low redshifts, and there is little chance that a single, merger-triggered 
burst of feedback can (alone) suppress all future growth. The division between the 
regimes appears 
to be at $\mhalo\sim10^{12}-10^{13}\,h^{-1}\,\msun$, 
interestingly similar to the traditional halo 
quenching mass (see \S~\ref{sec:quenching:maintenance:hothalo} below). 
We note that this analysis can be repeated in terms of the post-shock entropy, 
following \citet{scannapieco:sam}, which yields a nearly identical result. 

At the highest redshifts $z>2$, it is also difficult for a single event to suppress the cooling of 
all gas which will be accreted by these halos, especially for systems which are already 
massive at these redshifts (and therefore likely to form the most massive clusters at 
$z=0$). A $\sim10^{12}\,h^{-1}\,\msun$ halo at $z=4$, for example, is likely to 
grow to a $\sim10^{15}\,h^{-1}\,\msun$ cluster by $z=0$, so the baryon content contributing 
to the merger-driven feedback event at these redshifts is negligible compared 
to that which will be accreted at later times. 
%
%
We also caution that the feedback from a merger may not be as efficient 
as we have assumed in this analysis. Although we adopted a relatively 
conservative total ``stellar+quasar'' feedback energy input, it is not entirely 
clear how successful such feedback is at coupling to gas on large scales. 
Perhaps more important, the simple scalings above  
ignore the possibility that cooling instabilities 
might occur within the post-shock compression, or that cold clumps might 
be able to self-shield against a propagating shock, leaving most of the 
mass which would be accreted unaffected. 

In particular, if gas accretion 
occurs preferentially along filamentary structures, it may be difficult for 
feedback to directly couple to most of the gas in the filament.
A more detailed calculation of these effects will, unfortunately, require better knowledge of 
the actual drivers of feedback, as well as high-resolution 
simulations which can self-consistently resolve phase structure and shocks 
in the IGM gas. For now, we would more cautiously describe our calculations as 
estimates of what feedback from a major merger {\em could} do to suppress cooling. 
Even in this case, however, both our 
SPH simulations and simple scaling arguments 
suggest that the most massive systems, especially if their 
mergers occur early at $z\gtrsim2$, cannot be quenched just by the 
energy injection from a single feedback event.

\subsubsection{A ``Mixed'' Solution: Hot Halos from Quasar/Starburst Feedback}
\label{sec:quenching:maintenance:hothalo}

Given the uncertainties and limits, in the most massive systems, on the efficiency 
of short-term feedback from a major merger, we propose a mixed 
solution. Halos more massive than (roughly) $\sim10^{12}\,h^{-1}\,\msun$ 
have characteristic gas cooling timescales longer than the dynamical or free-fall timescale, 
and are so described as being in the so-called ``quasi-static'' or hot halo regime. 
In most models, ``radio mode'' feedback, i.e.\ {\em some} form of feedback from 
{\em low} accretion rate activity in a massive central BH provides the 
small additional heating term needed to maintain a pressure-supported 
hydrostatic equilibrium structure at all radii, preventing new gas from 
cooling onto the central galaxy. The additional heating term, for our purposes, does 
not even necessarily need to come from a central BH -- it could owe to 
kinematic heating or other effects, so long 
as it maintains the hot halo -- although energetic arguments \citep[e.g.][]{benson:sam} 
and high-resolution observations \citep[e.g.][]{batcheldor:outflow.mechanism} 
favor an AGN origin. 

A significant problem with these models, however, as we have seen in 
the \citet{croton:sam} example in \S~\ref{sec:ellipticals:fractions}, is that they are unable to 
produce sufficient numbers of red central galaxies in relatively low mass 
halos, and the red fraction does not depend on stellar mass as is observed. 
In other words, {\em some} process, with a dependence on galaxy mass, 
is required to assist the quenching of 
lower-halo mass systems. One might attempt to address this by adding a 
strong secular quenching mechanism, but we have shown 
by including this in the \citet{bower:sam} models that 
this fares little better at matching the bivariate red 
fraction as a function of halo and galaxy stellar mass, and that it conflicts 
with constraints on pseudobulge populations. 

However, we have just shown that feedback from a major merger 
can shock-heat sufficient surrounding gas to quench systems below 
the traditional hot halo mass threshold ($\sim10^{12}-10^{13}\,\msun$) 
for substantial periods of time. We therefore propose that traditional modes of 
quenching and feedback in hot halos remain the key to suppressing 
star formation in massive systems, but that these are supplemented 
by mergers, which can effectively quench star formation in lower-mass systems 
before these cross the hot halo threshold.
In fact, the major merger needs to suppress star formation in 
low mass systems only until they would naturally develop hot halos -- often 
much less than a Hubble time. 
For example, a typical $\sim10^{11}\,h^{-1}\,\msun$ halo merging at $z\sim4$ need only be 
quenched by merger feedback until $z\sim2$ ($\approx1.8\,$Gyr), when it 
will be sufficiently massive to enter the traditional hot halo regime. 
Once a hot halo is developed, the merger remnant already, by definition, 
has the means to maintain that halo and supplement it with feedback -- 
namely, a relatively massive spheroid and BH which will be accreting 
at low rates (i.e.\ the ideal seed for ``radio-mode'' feedback). 

More conservatively, the ``merger feedback'' does not even need to 
completely suppress cooling/accretion in these low mass systems. If the hot halo 
is an effective means of quenching, then mergers only have to create hot halos. 
In fact, the traditional hot halo is generated by an accretion shock in massive systems, 
and does not occur in low-mass systems because the conditions do not set up 
such a shock \citep{dekelbirnboim:mquench}. It is a small extension, then, to suppose that 
the strong shocks from merger-induced (quasar and starburst-driven) 
feedback, which are powerful even in 
low-mass systems, might accomplish this even when accretion shocks do not. 
Indeed, in Figure~\ref{fig:hot.halos} we show the cooling time relative to the 
free fall time ($t_{\rm ff}$) for the gas in our SPH merger remnants, and the amount of 
gas mass raised by the feedback coupling above a given $t_{\rm cool}/t_{\rm ff}$. 
Regardless of the mass of the systems or absolute values of the cooling times, 
the gas out to many times the virial radius is almost uniformly raised to 
$t_{\rm cool}/t_{\rm ff}\gg 1$, the traditional criterion for a hot halo. 

In other words, relatively low-mass halos (which would otherwise rapidly cool) 
require some event to enable their quenching and transition to a stable 
hot accretion mode (i.e.\ suppression of future cooling). We demonstrated in 
\S~\ref{sec:quenching:maintenance:nuke} that feedback associated with a 
major merger can easily accomplish this (although we note that feedback 
may be inefficient in extremely low mass halos, $\lesssim10^{10}\,\msun$, 
which are not important for our conclusions). 
Once a halo grows to large 
masses, however, any single quasar or stellar feedback event 
(or any baryonic feedback event, given the relative mass growth involved) 
is probably insufficient to 
singlehandedly heat the (very large) quantities of gas involved to a temperature 
so that the cooling time is longer than a Hubble time. However, in this regime, massive 
halos are already heating most of the gas via accretion shocks. Some 
mechanism (such as a major merger) is still needed to exhaust the gas in the 
central galaxy, and additional mechanisms (such as radio-mode feedback; 
requiring a massive spheroid and black hole in the remnant) may be 
needed to account for a small additional energy input or mixing term in the center 
of the halo (in order to prevent the formation of cooling flows at late times), but the 
bulk of the energetic input needed to maintain a hot system is already in place. 

We specifically check this scenario by revisiting 
Figure~\ref{fig:blastwaves}, and considering, instead of the 
amount of gas which can be heated to temperatures above which the 
cooling time is much longer than a Hubble time, the 
amount of gas which can be shocked to temperatures which are above 
the critical shock stability threshold; i.e.\ for which 
the cooling time is longer than the free-fall or dynamical time of the gas. 
Following \citet{dekelbirnboim:mquench}, we estimate this critical temperature  
for gas near the virial radius of the halos of interest, and obtain the 
(halo-mass independent) threshold 
$T_{\rm crit} \approx 4\times10^{5}\,{\rm K}\,\Delta_{200}^{1/2}\,(1+z)^{3/4}$. 
Interestingly, the ratio of BH to host mass (and therefore the 
relative amount of feedback energy coupled to the gas) appears to scale with 
redshift in a roughly similar manner \citep[see \paperone\ and][]{hopkins:bhfp}, yielding a 
nearly redshift-independent 
ratio of mass which can be shocked by merger feedback to that inside 
the virial radius at each epoch. In other words, at all redshifts, feedback 
from quasar and/or starburst activity associated with 
a major merger is sufficient to shock the entire gas content within the 
virial radius (or even to several times the virial radius) to this critical temperature, 
for halo masses $\mhalo\lesssim10^{13}\,\msun$. At larger halo masses, 
systems will already have naturally developed hot halos owing to 
accretion shocks, so it does not matter whether or not the 
feedback energy can shock the systems into the hot halo mode (although 
the merger-driven exhaustion and feedback may still be critical to ceasing 
star formation and making the system red).
And once the hot halo mode is 
established in the inner radii inside $R_{\rm vir}$, 
it does not ultimately matter how far the hot halo extends beyond 
the virial radius (or how much of the mass to be later accreted is affected) -- the hot halo 
sets up a quasi-static, pressure supported equilibrium against which newly 
accreted gas will shock and add to at large radii (regardless of its mass). 

A more detailed study of these hot halos from major merger-driven 
quasar and starburst feedback in 
cosmological simulations is an important topic of future work. However, it 
is ultimately a relatively small variation on the traditional principle 
which has been recognized for many years \citep[see][]{rees.ostriker.77,
norman.silk:gas.halos,blumenthal.84}. We 
have further shown that it is not only possible, but quite easily accomplished 
from moderate feedback prescriptions. 
Quenching can therefore be accomplished in the ``traditional'' context 
of hot halos supplemented by feedback from a massive BH, but 
allowing for feedback from black hole growth and 
star formation in a major merger, in a halo of {\em any} mass, to 
create a hot halo environment.

\section{Discussion}
\label{sec:discussion}

We have developed and tested a 
simple but physically-motivated model in order to study the cosmological 
role of mergers in the formation and quenching of red, early-type galaxies. 
By combining theoretically well-constrained 
halo and subhalo mass functions as a function of redshift and 
environment with empirical halo occupation models, we can 
predict the distribution of 
mergers as a function of redshift, environment, and physical galaxy properties. 
In \paperone, we discuss this methodology in detail, and show that it 
accurately reproduces a variety of observations over a 
wide range in redshifts, including 
observed merger mass functions; merger fractions as a function of 
galaxy mass, halo mass, and redshift; the mass flux/mass density in 
mergers; the large-scale clustering/bias of merger populations; 
and the small-scale environments of mergers. 
The primary advantage of this model is that it allows us to 
study and make a priori predictions for 
the effects of mergers without many of the uncertainties or 
degeneracies inherent in present cosmological simulations or semi-analytic models. 

For example, cosmological simulations still lack the resolution to 
model the processes of internal galactic kinematics in mergers, black 
hole accretion/feedback, and disk formation. Although progress is being 
made studying these processes via ``zoom-in'' simulations, it is not meaningful to 
speak of gas-rich, spheroid forming mergers in cosmological 
populations if a cosmological box does not contain the appropriate, representative 
population of accurately formed disk galaxies (the progenitors 
in these mergers) in the first place.

Although semi-analytic models 
avoid some of these difficulties, they require making a number of assumptions 
regarding models or physics that we are not attempting to 
test in this paper, including e.g.\ 
disk formation, star formation efficiency in disks, disk instabilities, 
minor mergers, satellite disruption, reddening of satellite galaxies, 
and the exact physical mechanisms of feedback. These assumptions 
introduce uncertainties in the model and, more importantly, obscure the 
key physical elements being tested. 

Our adopted model, in contrast, 
bypasses these (unnecessary for our purposes) assumptions and 
uncertainties, and instead 
empirically adopts the relevant consequences of all these physical 
processes -- namely what kinds of galaxies are merging at a given place and time. 
We can then more directly ask the question we wish to answer: 
how do mergers contribute to the formation and/or quenching of massive 
red galaxies? 

We find that the simple assumption that star formation is quenched after a gas-rich, 
spheroid-forming major merger (by any mechanism) 
naturally predicts the turnover in the galaxy mass-halo mass relation 
at $\sim\lstar$ -- i.e.\ the fundamental turnover in the efficiency of star formation 
and incorporation of baryons in galaxies, at the observed scale and without 
any parameters tuned to this value. The physical scale $\sim\lstar$ reflects 
the point where major, galaxy-galaxy mergers first become efficient. At lower 
masses, major mergers are rare -- this is true both of 
halo-halo major mergers \citep[e.g.][]{vandenbosch:subhalo.mf} 
and galaxy-galaxy mergers (which 
are further suppressed at these masses because of the relative 
scalings of orbital velocities and internal galaxy velocities -- i.e.\ 
two such galaxies are likely to interact as field flyby or satellite-satellite 
systems with relatively high orbital velocities that do not efficiently merge). 

Systems therefore generally grow uninterrupted, potentially building 
relatively low-mass pseudobulges ($\lesssim10^{10}\,\msun$) via disk/bar instabilities 
or minor mergers, until they get to $\sim\lstar$. By these masses, the probability of 
the halo merging with a major companion reaches of order unity, and 
the velocity scalings are such that the two galaxies (once the halos have merged) 
will merge efficiently ($t_{\rm merger}\ll \tH$). The systems 
can then grow via subsequent (dry) mergers, but this is a relatively inefficient channel 
(i.e.\ mass growth is slow). 
Because their star formation is quenched (and therefore no longer 
keeping pace with their host halo growth), mergers themselves also rapidly become less efficient 
(i.e.\ the system mass becomes low relative to the host halo mass, increasing 
the merger timescales). 

In addition, our model naturally predicts the 
observed mass functions and mass density of red galaxies as a function of 
redshift, the formation times of early-type galaxies as a function of mass, 
the fraction of quenched galaxies as a function of galaxy and halo mass, environment, 
and redshift, and the distribution/dichotomy of kinematics in massive ellipticals. 
Each of these predictions agrees well with observations over the 
entire observed range of galaxy masses and redshifts. 
As demonstrated in \paperone, our 
model also agrees well with observed merger rates and fractions as a 
function of galaxy mass and halo mass at all observed redshifts. Together with 
the agreement between our model and the observed mass functions and mass density 
of red galaxies, this illustrates that there are, in fact, sufficient numbers of mergers 
(both in theory and observed) to produce the entire massive 
spheroid population at all observed redshifts \citep[see also][]{hopkins:transition.mass}.
Also, unlike commonly adopted models in which quenching is regulated purely 
by halo mass, we have not adjusted or tuned any parameter to 
give the desired results. Indeed, there is not even an obvious parameter 
which can be tuned to give the turnover in the galaxy mass-halo mass 
relation at the appropriate location (since it appears not to depend  
on our calculation of the merger timescale). To the extent that mergers 
can supplement quenching, then, this suggests that it is not necessarily 
problematic that theoretical calculations \citep{birnboim:mquench,keres:hot.halos} 
do not give exactly the same 
halo quenching threshold as semi-analytic models subsequently tuned 
to fit the observations, as has been noticed in several works 
\citep[e.g.][]{croton:sam,cattaneo:sam}. 

Although these predictions are suggestive, recent semi-analytic models 
have demonstrated that many of them are non-unique. A variety of 
different quenching implementations and feedback effects in these 
models have been shown to successfully reproduce e.g.\ low-redshift 
mass functions, color-magnitude diagrams, 
and mean red fractions. We therefore investigate the robust, observable 
differences between three broad classes of models for quenching. 

First, our adopted merger-induced quenching model, in which some 
mechanism enables merger remnants to remain quenched. 
Second, a halo quenching model, in which quenching is primarily 
determined by a simple (albeit potentially redshift-dependent) 
halo mass threshold (regardless of merger history 
or morphology) -- i.e.\ one in which some mechanism enables any system to 
remain quenched if and only if it develops a ``hot halo.'' Third, a 
secular model, in which color (and/or morphological) transformation is 
driven solely by galaxy structure (essentially baryonic galaxy mass), 
owing to e.g.\ disk or bar instabilities (or other non-merger related mechanisms). 
Regardless of the exact details of their quenching prescriptions (and other 
assumptions), most present semi-analytic models can clearly be 
identified with one of these three classes of models, based on which 
criterion effectively dominates quenching (e.g.\ galaxy merger history, halo mass, 
or disk mass), and we demonstrate that the key qualitative predictions of 
each class will remain true. Note that we are explicitly referring to the 
quenching of {\em central} halo galaxies (the great majority of 
$\gtrsim10^{10}\,\msun$ galaxies), as the reddening of satellites is almost 
certainly affected by other processes (such as their initial 
accretion, ram pressure stripping, or harassment). 

We show that these models make a number of robust, unique predictions
with respect to several observables, including: 

{\em (1) Bivariate Red Fractions:} Observational measurements of the 
red fractions of galaxies in groups can now break the degeneracy between 
the fraction of quenched systems as a function of galaxy mass (which 
all these models successfully reproduce) and halo mass. The observations 
show several important qualitative trends in the fraction of quenched, 
central halo galaxies as a bivariate function of galaxy stellar and halo 
mass \citep[e.g.][]{weinmann:obs.hod}. 
These include: (1) a strong dependence of red fraction on halo mass,  
(2) some (weaker) residual dependence on galaxy mass/luminosity, (3) 
a lack of any sharp characteristic scale in $\mhalo$, (4) a relatively high 
red fraction ($f_{\rm red}\gtrsim0.5$) for the most massive/luminous systems even 
at relatively low halo masses ($\mhalo\lesssim10^{12}\,\msun$), and 
(5) a similar, relatively high red fraction ($f_{\rm red}\gtrsim0.5$) 
for the least massive/luminous systems at high halo masses 
($\mhalo\gtrsim10^{13}\,\msun$).
The fundamental difference between the classes of models we consider 
is directly reflected in this predicted bivariate red fraction (where 
we refer specifically to central galaxies, as satellites may be affected by 
other processes as indicated above). 

In halo models, the red fraction is essentially a step function in halo mass 
with a sharp transition from low red fractions to 
$f_{\rm red}\sim1$ 
around the critical quenching mass, and little residual dependence on 
galaxy properties. In secular models, the red fraction is just a function 
of galaxy mass, with little (or even inverse, if quenching becomes 
harder to maintain in high mass halos) correlation with halo mass. 
Mergers, however, depend on both galaxy and halo mass, with 
larger galaxies at a given halo mass merging more efficiently 
(and being more likely to have already undergone a major merger), 
while larger halos are more evolved and more likely 
to have accreted a major companion as fuel for a major merger. 
Consequently, the red fraction is an increasing function of halo mass, but 
with an additional (weaker) dependence on galaxy mass, and grows smoothly 
(i.e.\ without a single, sharp characteristic scale) to higher masses. 
A significant dependence on halo mass is maintained, but there 
is still a large red fraction for the massive galaxies (even in relatively 
low-mass halos). More detailed observations are needed to quantify this 
in greater detail, but only the merger model appears to match 
the qualitative trends observed.

{\em (2) High-Redshift Passive Galaxies:} A relatively large population of 
massive, red galaxies exists at even high redshifts $z\gtrsim3$. 
Although at high redshifts most (simply identified) ``red'' galaxies 
are dusty star-forming systems, there is a significant population 
which are truly ``red and dead,'' spectroscopically confirmed 
passively evolving, low star-formation rate spheroids 
\citep{labbe05:drgs,kriek:drg.seds}. In contrast, 
semi-analytic halo quenching models are generally forced to assume that 
some process (e.g.\ accretion in filaments or cold clumps) at these redshifts 
raises the mass threshold for quenching, and as a result the predicted 
density of passive galaxies drops rapidly at $z\gtrsim2$. 
We note that this is not a statement that ``hot halos'' cannot or 
do not form at these redshifts \citep[simulations, in fact, suggest that they 
do;][]{keres:hot.halos}, 
nor that such models do not predict a sufficient density of all massive 
galaxies at these redshifts. 
However, in the naive implementation (in which 
the quenching is strongly dominated by a simple halo mass threshold), 
one cannot simultaneously form massive galaxies (and predict a 
sufficiently high global star formation rate density) at high redshift {\em and} 
quench them. In order to match both the observed density of 
star-forming and passive massive galaxies, some mechanism is 
required which can explain the quenching of {\em some}, but not all, systems in 
massive halos at high redshifts. 

Mergers, on the other hand, proceed efficiently 
in massive halos at high redshifts, predicting a significant density of 
quenched, passively evolving systems even at $z\gtrsim3$, in 
good agreement with the observations. A secular model can also 
explain the density of passive systems at these redshifts, since, by definition, 
the existence of such massive galaxies in the first place guarantees that a large 
fraction will be red (since the red fraction is a pure function of galaxy mass 
in this model). However, the secular model encounters a different 
conflict at high redshift.

{\em (3) Buildup of the Color-Density Relation:} The color-density 
relation appears to weaken with redshift, flattening in intermediate 
density environments until $z\sim1.5$, where it appears 
that there is no measurable color-density relation in field environments 
\citep{nuijten:color.density.evol,cooper:color.density.evol,gerke:blue.frac.evol}. 
Even at high redshifts $z\sim3$, however, 
there is still a significant color-density relation 
\citep{quadri:highz.color.density} -- it is simply that the relation becomes significant only in 
more extreme (proto-cluster, for example) environments. In other words, 
a large population of quenched galaxies 
emerges rapidly at early times in the 
most massive environments, and then subsequently builds up in more 
moderate environments at lower redshifts. 

A halo quenching model 
predicts something similar to the low-redshift evolution in these 
trends (although with difficulty in producing quenched systems in 
all but the most truly extreme environments at high redshift, as described above). 
At higher redshift, systems above the halo threshold quenching 
mass represent progressively more extreme environments, and 
if this effective mass threshold increases with redshift, the trend is more 
pronounced. The red fraction is still nearly a step-function at each redshift, 
but with a shifting relative scale. 

A merger model also predicts a 
trend qualitatively similar to that observed. The most dense environments 
undergo their epoch of major mergers more rapidly than less dense 
environments (equivalently, more massive present environments passed 
through their small group stage at earlier times), although 
the red fraction in halos of all masses decreases with redshift 
(as there is less time for mergers to operate). By $z\sim1.5$, typical 
field environments have uniformly low red fractions, and no significant 
measurable color-magnitude relation is expected. The location of 
the buildup of quenched galaxies shifts to denser environments, similar to the 
observed trend. 

A secular model, in contrast, predicts almost no evolution 
in the trend of red fraction with halo mass as a function of redshift (as a 
consequence of there being relatively little evolution in the average mass of 
a star-forming galaxy hosted by a given halo mass). If one allows for 
high-redshift disks being more compact, this increases their inferred instability, 
yielding {\em opposite} evolution in the red fraction versus halo mass 
to that observed (i.e.\ increasing quenched fractions with redshift at fixed $\mhalo$).
As a consequence, although there may be some artificial evolution with redshift in 
the red fraction as a function of environment (as the same halo mass corresponds to 
different environments), there is no significant true evolution. Furthermore, in 
a secular model, the 
halo masses corresponding to field environments do not trend towards uniformly 
low red fractions by $z\gtrsim1.5$ -- i.e.\ there is little ``smearing out'' of the 
color-density relations at high redshift. Future observations are needed to 
make these comparisons formal, but quantifying the evolution with redshift in 
the red fraction as a function of host halo mass (from large samples which 
can isolate groups and group central galaxies, and span a wide range of environments) 
will be a powerful discriminant between these models.

{\em (4) Spheroid Kinematics (Dichotomy of Elliptical Galaxies):} Numerical 
simulations and observations of merger remnants and elliptical kinematics 
demonstrate that gas-rich major mergers 
(i.e.\ those involving disks, even with low gas fractions $\fgas\lesssim0.1$) 
generally produce typical low-mass ($\lesssim$ a few $\lstar$) 
ellipticals with central cusps, disky isophotes, and significant rotation, 
while subsequent gas-poor spheroid-spheroid mergers produce 
typical high-mass ellipticals with central cores, boxy isophotes, 
and little rotation. There is a well-defined transition between the two 
classes of spheroids, at a mass $\sim2-3\times10^{11}\,\msun$ for 
each of these criteria. 

A merger model naturally predicts this transition 
point: at lower mass, most spheroids (i.e.\ merger remnants) have 
experienced only their initial, disk-disk spheroid-forming merger or 
(in some cases) one additional, gas-rich, disk-spheroid major merger. 
At higher mass, most systems have undergone an additional, subsequent 
spheroid-spheroid major merger. 
If the major merger is associated with quenching, the low-mass
disk mergers are guaranteed to 
be gas-rich, and the spheroid-spheroid mergers at high masses 
are guaranteed to be gas-poor, matching the observed trends and transition point 
in each of the cusp/core, disky/boxy, rapid/slow rotation criteria. 

In a halo quenching model, however, many systems undergo their first 
major merger somewhat before their host halos cross the quenching 
mass threshold, and therefore re-accrete significant disks. Their subsequent 
mergers are not gas-poor, spheroid-spheroid any longer, but 
gas-rich, disk mergers. As a result, the predicted transition mass between disky 
and boxy ellipticals is increased by an order of magnitude (only the most massive 
cD galaxies cross the quenching mass threshold early enough to have 
had multiple subsequent major mergers since that time), in contradiction 
to the observations. 

A secular model suffers from the opposite problem. In order 
for secular mechanisms to dominate quenching, they must act before major mergers 
transform the system to a spheroid -- i.e.\ systems must (by definition in such a model) 
predominantly quench before they undergo their first merger. A large fraction of 
even these first mergers, then, are gas-poor, spheroid-spheroid 
(or pseudobulge-pseudobulge) mergers. The predicted transition mass between 
disky and boxy ellipticals is therefore decreased by an order of magnitude, 
again in contradiction to the observations. It appears that matching 
the observed transition in elliptical types (without violating the basic 
kinematic constraints from simulations and observations) fundamentally requires 
some quenching of massive spheroids/merger remnants \citep[a conclusion also reached 
by][who begin from a halo quenching model]{naab:dry.mergers}.

{\em (5) Effects of Small-Scale Environment:} It has also been 
suggested observationally that the red fraction (at fixed halo mass and 
galaxy mass) does not depend on large-scale environment, but 
may depend on small-scale environment, in the sense that it increases 
with overdensities on small scales \citep{blanton:smallscale.env}. 
We caution that, at present, interpretation of these observations is 
difficult because they include both satellite and central galaxies. However, 
if the result is borne out for central galaxies alone (i.e.\ a central galaxy 
is more likely to be red, all else being equal, if it lives in a small-scale 
overdensity) via measurements of the cross-correlation function for 
red, central galaxies and other galaxies, then this would also favor a merger model. 
We note that it is not necessary that merger remnants live in such overdensities 
long after their mergers (as, by definition, the mergers will consume 
some of the very galaxies that define such an overdensity). However, 
if such a trend exists, it is difficult to explain in a pure halo quenching or 
secular model, as both mechanisms operate independent of 
neighboring galaxy populations. 

These consequences of merger-driven, halo mass-driven, and secular/disk instability-driven 
quenching models are robust, and future observations should be able to
break the degeneracies between the models. Although the quantitative details 
may differ slightly in different implementations of the models, we have shown 
that current state-of-the-art semi-analytic models \citep[e.g.][which include a number of 
other prescriptions and more detailed physical recipes than the 
toy models described above]{croton:sam,cattaneo:sam,bower:sam} 
fundamentally yield predictions which are qualitatively 
identical to the behavior expected for the basic classes of models described above. 
These behaviors are generic to any model in which these processes dominate 
the quenching of central galaxies, and the predictions shown are at least qualitatively 
robust regardless of ``tuning'' the models. 

That we have not tuned or adjusted our model to give a particular result should not, 
of course, be taken to mean that there are no uncertainties in our approach. 
However, we 
re-calculate all of our predictions adopting different estimates for the 
subhalo mass functions and halo occupation model (and its redshift 
evolution) and find this makes little difference (a factor $<2$) at all 
redshifts. The largest uncertainty comes from our calculation of 
merger timescales, where, at the highest redshifts ($z\gtrsim3$), merging via 
direct collisional processes may be more efficient than 
merging via dynamical friction, given the large physical densities. 
More detailed study in very high-resolution numerical simulations will 
be necessary to determine the effective breakdown between different 
merger processes. 
Nevertheless, the difference in our predictions at these redshifts is still 
within the range of observational uncertainty. 

Ultimately, we find that our predictions are robust 
above masses $\mgal\gtrsim10^{10}\,\msun$, regardless of these 
changes to our model, as the theoretical 
subhalo mass functions and empirical halo occupation models 
are reasonably well-constrained in this regime. Below these masses, 
in any case, 
it is likely that a large fraction of spheroids are relatively small 
bulges in disk-dominated galaxies (of which a large fraction 
may be pseudobulges formed by disk instabilities) and that a large fraction of 
the red galaxy population are satellites (whose reddening 
may be affected by their mere accretion as a satellite, let alone tidal 
or ram pressure stripping processes, which we do not attempt to model). 
While not dominant in the $\gtrsim\lstar$ galaxies with which our modeling 
is most concerned, these processes are certainly important for 
low-mass populations. 

We further discuss a variety of physical mechanisms that 
may drive the quenching of major merger remnants. 
In numerical experiments, the star formation rates of isolated disks 
(i.e.\ ones cut off from any gas accretion) decay slowly, and the galaxies do not 
move to the red sequence in times $\lesssim$\,a few Gyr (despite allowing 
for secular instabilities in these simulations). This alone is a consideration which 
should be of concern in secular or pure halo quenching models -- without 
mergers or some other driver of violence in the system, these systems 
do not efficiently transition to the red sequence in the first place. 
However, it is clear that 
merger remnants efficiently exhaust gas and redden rapidly onto 
the red sequence, even 
without the inclusion of feedback effects (although these may be necessary 
to fully terminate star formation in the most high-redshift, gas-rich systems). 
It is clear that mergers easily accomplish the ``transition'' to the red sequence, 
even if only temporarily. The more difficult question is how such systems 
might prevent future cooling, in order to remain quenched for significant 
periods of time.

There are, however, a number of feedback sources directly associated 
with major mergers, including purely kinematic ``stirring,'' tidal heating, and 
shock effects, long-lived starburst-driven winds, and (potentially) 
impulsive, quasar-driven outflows. We demonstrate in numerical simulations 
(and from simple scaling arguments) that the combination of these 
feedback effects (even with relatively mild prescriptions for their 
strength) is sufficient to heat several times the initial baryon content of the 
host halo at the time of the merger to very high temperatures, 
at which the cooling time becomes longer than a Hubble time. For 
the $\sim1/2$ of the present $\sim\lstar$ red galaxy population that has moved 
onto the red sequence since $z\sim1$, this single feedback event is sufficient to 
prevent all but $\lesssim1\%$ of the galaxy mass from cooling back onto 
the galaxy by $z=0$, i.e.\ sufficient to ensure the galaxy remains ``red and dead.'' 
The problem, however, is potentially more severe at high redshifts. 
Not only must the suppression of cooling act for a longer period of time, 
but a massive halo at high redshifts $z\gtrsim2$ may typically grow by a large 
amount (more than an order of magnitude) in mass by $z=0$, implying that 
the total baryon content for which cooling must be suppressed is larger 
than that of the galaxy. Moreover, the increased densities at these times 
further suppress the propagation of feedback-driven shocks, and 
enhances the cooling rates by large factors ($\sim100$ at $z\sim3-4$). 

We therefore propose a ``mixed'' solution, in which 
traditional modes of 
quenching and feedback in quasi-static ``hot halos'' remain the key to suppressing 
star formation in massive systems, but that these are supplemented 
by mergers, which can effectively quench star formation in lower-mass systems 
before these cross the hot halo threshold. The major merger 
can supplement this traditional quenching mode in two ways: first, by temporarily 
suppressing cooling until the system naturally develops a hot halo (i.e.\ crosses 
the quenching mass threshold). This is, even for high redshift systems, 
often much less than a Hubble time ($\sim$a couple Gyr), and is relatively 
easily accomplished by the feedback effects described above. 
Second, the strong shocks from the merger-driven feedback can accomplish 
what an accretion shock would in a more massive system -- i.e.\ they can 
create a quasi-static hot halo even in low-mass systems which would not 
(independent of a major merger) develop such a halo on their own. 

We demonstrate both using
simple scaling arguments and numerical simulations including 
feedback, cooling, star formation, and realistic shock mechanisms, 
that even conservative feedback prescriptions will 
shock most of the gas within the virial radius to temperatures and entropies 
where the cooling time becomes much longer than the free fall or dynamical/compression 
timescales (the traditional definition of a hot halo). 
Once this hot halo is established inside $R_{\rm vir}$, 
a quasi-static, pressure supported equilibrium is established against which newly 
accreted gas will shock and add to at large radii (regardless of the mass 
subsequently accreted). 
The energetics of merger-triggered 
feedback are sufficient to achieve this in all halos $\mhalo\lesssim10^{13}\,\msun$ 
(including halos below the traditional quenching mass threshold), 
with little dependence on redshift (at least from $z=0-6$). 
Once a hot halo is developed, the problem of maintaining that hot halo 
(i.e.\ preventing cooling flows) is no different from the traditional cooling flow problem 
(which we are not directly attempting to address here), but 
the merger remnant already, by definition, 
has the means to heat the halo and supplement it with feedback -- 
namely, a relatively massive spheroid and BH which will be accreting 
at low rates (i.e.\ the ideal seed for ``radio-mode'' feedback). 

This ultimately simple variation on the traditional models of quenching 
in massive systems appears to yield a number of qualitatively different 
predictions, as described above, and merits further study. 
Although, in order to limit the physical assumptions being studied, we 
did not adopt a full semi-analytic model, it will be valuable for 
future studies and comparison to observations to implement 
such models.
There are a number of prescriptions one might consider, with 
varying degrees of complexity, 
which may yield different, testable observational predictions. 
Ideally, such models should consider a variety of prescriptions for 
quenching, and compare the results in order to determine 
what (if any) observational tests might break the degeneracies 
between them.

(1) Pure Merger Quenching: This is the simplest possible model, 
similar to what we have assumed in this work, assuming that a major 
merger completely suppresses future cooling/star formation. Equivalently, 
one could adopt some bulge-to-disk ratio above which cooling is 
suppressed, as in \citep{cattaneo:sam}, or a bulge mass threshold 
\citep[$M_{\rm bulge}\gtrsim3\times10^{10}\,\msun$, as in][]{naab:dry.mergers}. 

(2) Merger Feedback/Strong Shocks: Rather than fiat quenching, 
one could allow for some large energy injection from feedback (presumably owing 
to triggered quasar and starburst activity)
in a merger, and assume that the appropriate shocked quantity of 
gas has its cooling suppressed, or is ejected from the host galaxy and 
reheated to the halo virial temperature \citep{somerville:new.sam}. 
This is similar to the calculation 
in \citet{scannapieco:sam}, who demonstrate that such an assumption is 
sufficient to produce downsizing trends below $z\sim2$. 
There are a number of analytic models which have been proposed for 
the effects of this feedback, including the blastwave model calibrated 
to simulations in \citet{hopkins:seyferts}, the 
model of \citet{scannapieco:sam} in terms of the post-shock entropy, 
and the temperature/cooling time calculations in \S~\ref{sec:quenching:maintenance}. 

(3) Merger-Induced ``Hot Halos'': Based on the arguments above, 
it is straightforward to assume that feedback from 
quasar and/or starburst activity triggered in a major merger drives the host 
halo to the quasi-static, hot halo regime. Whatever the treatment in the 
semi-analytic model is for such hot halos (i.e.\ whether they are
assumed to be quenched, or whether various AGN feedback modes 
are considered for ``maintenance'' purposes), the host halos of major 
merger remnants would be treated identically. 

(4) A ``Full Model'': Ideally, semi-analytic models could incorporate all of the 
effects above. Based on energetic arguments or the simple scaling 
arguments in \S~\ref{sec:quenching:maintenance}, or adopting some analytic model for 
a feedback-driven shock \citep{scannapieco:sam,hopkins:seyferts}, 
one can calculate the appropriate effects on 
halo gas. It is then possible to consider whether this moves the halo 
into the hot halo regime, or ``buys time'' until the halo experiences 
accretion shocks and falls into such a regime itself. Feedback from low-luminosity 
AGN, or cyclic accretion inside a hot halo, would be allowed, and could 
further suppress subsequent cooling. 

Future study using high-resolution numerical simulations will be essential to 
ultimately understanding the interplay of these complex feedback 
processes. Simulations with the dynamic range to 
simultaneously resolve the relevant galactic structure and feedback processes 
and cosmologically rare, massive populations are 
not yet feasible; however, the effects of these processes in 
representative systems can be studied in detailed zoom-in simulations
\citep{li:z6.quasar}. 
Examining, for example, the effects of feedback on clumpy accretion 
at high redshift or the details of how merger-driven shocks transform the 
halo cooling structure will be critical to inform theoretical models of 
how these systems quench and suppress cooling over cosmic time. 
The combination of detailed simulations used to study the effects of 
feedback and cosmological models which enable predictions for 
the broad statistical properties of rare populations 
should allow future observations to break the 
degeneracies between different quenching models and 
tightly constrain the history of massive galaxy formation. 

\acknowledgments We thank Marijn Franx, Rachel Somerville, Richard Bower, 
Michael Cooper, Thorsten Naab, Ivo Labb{\'e}, and Norm Murray 
for helpful discussions contributing to this paper. 
This work was supported in part by NSF grant AST
03-07690, and NASA ATP grants NAG5-12140, NAG5-13292, and NAG5-13381.

\bibliography{ms}

\end{document}